\documentclass[aps,pre,superscriptaddress,twocolumn]{revtex4}
\usepackage{amsfonts,amssymb,amsmath,latexsym,epsfig,wasysym,graphics,graphicx} 
\usepackage[sort&compress]{natbib}

\usepackage{lscape} 
\usepackage{longtable} 
\usepackage{setspace} 
\usepackage{rotating}
\usepackage{hyperref}

\newcommand{\bx}{\bar{x}}

\def\sm{\sigma_\tau/\langle \tau \rangle}

\def\sm{\sigma_\tau/\langle \tau \rangle}

\begin{document}

\title{On the origin of long-range correlations in texts}

\author{Eduardo G. Altmann} 
\affiliation{Max Planck Institute for the Physics of Complex Systems, 01187 Dresden, Germany}

\author{Giampaolo Cristadoro}
\affiliation{Dipartimento di Matematica, Universit\`a di Bologna, 40126 Bologna, Italy}
\author{Mirko Degli Esposti}
\affiliation{Dipartimento di Matematica, Universit\`a di Bologna, 40126 Bologna, Italy}


\begin{abstract}

The complexity of human interactions with social and natural phenomena is mirrored in the way we describe our experiences through natural
language. In order to retain and convey such a high dimensional information, the statistical properties of our linguistic output has to be
highly correlated in time.  An example 
are the 
robust observations, still largely not understood, 
of correlations  on arbitrary long scales in literary texts. In this
paper we explain how long-range correlations flow from highly structured linguistic levels down to the 
building blocks of a text (words, letters, etc..). By combining calculations and data analysis we show that correlations take form of a
bursty sequence of events once we approach the semantically relevant topics of the text. The mechanisms we identify 
are fairly general and can be equally applied to other hierarchical settings.  \\
Published as: \href{dx.doi.org/10.1073/pnas.1117723109}{Proc. Nat. Acad. Sci. USA (2012) doi: 10.1073/pnas.1117723109}
\end{abstract}

\keywords{long correlations | complex systems | language dynamics | statistical physics | burstiness}
\maketitle


Literary texts are an expression of the natural language ability to project complex and
high-dimensional  phenomena into a one-dimensional, semantically meaningful sequence of symbols. For this projection to be successful, such
sequences have to encode the information in form of structured patterns, 
such as
correlations on arbitrarily long
scales~\cite{schenkel,eckmann1}. 
Understanding how language processes long-range correlations, an ubiquitous signature of complexity 
present in human activities~\cite{vossMusic,gilden,bunde,rybski,kello} and in the natural world~\cite{press,li,peng,voss},
is an important task towards comprehending how natural language works and evolves.  
This understanding is also crucial to improve the increasingly important applications of information theory and statistical natural language 
processing, which are mostly based on short-range-correlations methods
 \cite{manni,stamata, ober, usatenko}.   

Take your favorite novel and consider the binary sequence obtained by mapping each  vowel into a $1$ and all other symbols into a
$0$. One can easily detect structures on neighboring bits, and we certainly expect some repetition patterns on the size of words. But one should
certainly be surprised and  intrigued when discovering that there are structures (or memory) after several pages or even on arbitrary large
scales of this binary sequence. 
In the last twenty years,  similar observations of long-range correlations in texts have been related to large scales 
characteristics of the novels 
such as the story being told, the style of the book,  the author, and  the language~\cite{schenkel, eckmann1,amit,ebeling1,ebeling3,allegrini, melnyk2005,montemurro}.
However, the mechanisms explaining these connections are still missing (see
Ref.~\cite{eckmann1} for a recent proposal). Without such mechanisms, many fundamental questions cannot be answered.
For instance, why all previous investigations observed long-range correlations despite their radically different approaches? How and which correlations
can flow from the high-level semantic structures down to the crude symbolic sequence in the presence of so many arbitrary influences?  
What information is gained on the large structures by looking at smaller ones? Finally, what is the origin of the long-range correlations?

\begin{figure}[!ht] 
\includegraphics[width=1\columnwidth]{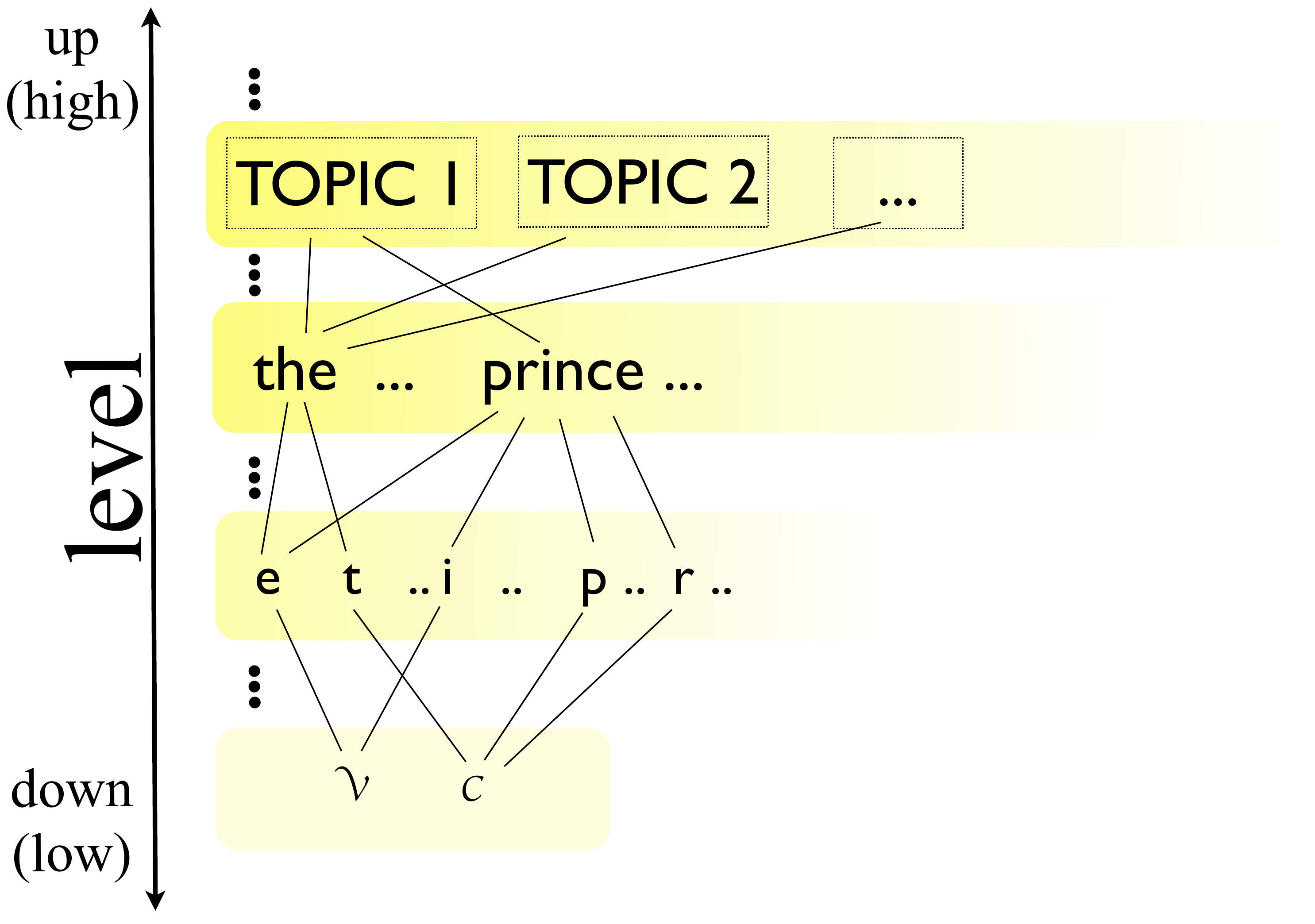}
\caption{ Hierarchy of levels at which literary texts can be analyzed.
Depicted are the levels vowels/consonants ($\mathcal{V}/\mathcal{C}$), letters (a-z), words, and topics. 
}\label{fig.1}
\end{figure}

In this paper we provide answers to these questions by approaching the problem through a novel
theoretical framework. This framework uses the hierarchical organization of natural language to identify a mechanism that links the
correlations at different linguistic levels. As schematically depicted in Fig.~\ref{fig.1}, a topic 
is linked to several words that are used to describe it in the novel. At the lower level, words are connected to the letters they are formed, and so on. We calculate how
correlations are transported through these different levels and compare the results with a detailed statistical analysis in ten different 
novels. Our results reveal that while approaching semantically
relevant
high-level structures, correlations   
unfold
in form of a bursty
signal. Moving down in levels, we show that correlations (but not burstiness) are preserved, explaining the
ubiquitous appearance of long-range correlations in texts.

\section{Theoretical framework}\label{sec.theory}
\subsection{The importance of the observable}\label{ssec.encoding}
In line with information theory, we treat a literary text as the
output of a stationary and ergodic source  
that takes values in a finite alphabet and we look for information
about the source through a statistical analysis of the text~\cite{cover}. 
Here we focus on correlations functions, which are defined after specifying an observable and a product over functions. 
In particular, given a symbolic sequence \textbf{s} (the
text), we denote by $s_k$ the symbol in the $k$-th position and  by $s_n^m$ ($m \ge n$) the substring $(s_n,s_{n+1},\ldots,s_m)$. As observables, we  consider  
functions $f$ that map symbolic sequences \textbf{s} into a sequence  \textbf{x} of  numbers  (e.g., $0$'s and $1$'s). We
restrict to local  mappings, namely $x_k=f(s_k^{k+r})$  for any $k$ and a finite constant $r \geq 0$. Its  autocorrelation
function is defined as: 
\begin{equation}\label{eq.2point}
C_f(t):=\langle  f(s_i^{i+r})f(s_{i+t}^{i+t+r}) \rangle - \langle f(s_{i}^{i+r})\rangle \langle f(s_{i+t}^{i+t+r}) \rangle,
\end{equation}
where $t$ plays the role of 
time (counted in number of symbols) and $\langle \cdot \rangle$ denotes an average over
sliding windows, see Supporting Information (SI) Sec. I for details.

The choice of the observable $f$ is crucial in determining whether and which
``memory'' of the source is being quantified.
Only once a class of observables sharing the same properties is shown to have the same asymptotic autocorrelation,
it is possible to think about  long-range correlations of the text as a whole.
In the past, different kinds of observables and encodings (which also correspond to particular choices of $f$) were used,
from  the Huffmann code~\cite{grassberger}, 
to attributing to each symbol an arbitrary binary sequence (ASCII, unicode, 6-bit tables, dividing letters in groups,
etc.)~\cite{schenkel,amit,kokol,kanter,melnyk2005}, 
to the use of the frequency-rank~\cite{monte} or parts of speech~\cite{allegrini} on the level of words. 
While the observation of long-range correlations in all cases points towards a fundamental source, it remains unclear which common
properties these observables share.  
This is essential to determine whether they share a 
common root (conjectured in Ref.~\cite{schenkel}) and to understand the meaning of quantitative changes in the correlations for different
encodings (reported in Ref.~\cite{amit}).
In order to clarify these points we use mappings~$f$ that avoid the introduction of spurious correlations.
Inspired by Voss~\cite{voss} and Ebeling {\it et al.}~\cite{ebeling1,ebeling3}\footnote{Our approach is slightly different from Refs.~\cite{voss,ebeling1,ebeling3} because instead of performing an average over different symbols we investigate each symbol separately.}
we use $f_\alpha$'s that transform the text into binary sequences  $\textbf{x}$ by assigning $x_k=1$ if and only if a  local matching condition $\alpha$ is satisfied at the $k$-th symbol, and $x_k=0$
otherwise (e.g., $\alpha=$ {\it k-th symbol is a vowel}). See SI-Sec. II for specific examples. 

\subsection{Correlations and burstiness}\label{ssec.correl}

Once equipped with the binary sequence~${\bf x}$ associated with the chosen condition $\alpha$ we can investigate the asymptotic trend of 
its $C_\textbf{x}(t)$. We are  particularly interested in  the
long-range correlated case
\begin{equation}
C_{\bf x}(t):= \langle x_j x_{j+t} \rangle -\langle x_j \rangle\langle x_{j+t} \rangle \simeq t^{-\beta}, \qquad  0< \beta < 1,
\end{equation}
for which $\sum_{t=0}^{\infty} C(t)$ diverges.
In this case the 
associated random walker $X(t):=\sum_{j=0}^t x_j$ 
spreads super-diffusively as~\cite{voss,trefan}
\begin{equation}\label{eq.mu}
\sigma_X^2  (t) := \langle X(t)^2 \rangle -  \langle X(t) \rangle^2 \simeq t^{\gamma}, \qquad \gamma=2-\beta.
\end{equation}

In the following we investigate 
correlations of the binary sequence~${\bf x}$ using Eq.~(\ref{eq.mu}) because
integrated indicators lead to more robust numerical estimations of asymptotic
quantities~\cite{peng,schenkel,voss,ebeling1}. We are mostly
interested in the distinction between short- $(\beta>1,\gamma=1)$ and long- $(0<\beta<1,1<\gamma<2)$ range correlations. We use
normal (anomalous) diffusion of $X$ interchangeably with short- (long-) range correlations of~\textbf{x}.

An insightful view on the possible origins of the long-range correlations can be achieved by exploring the relation between the power
spectrum $S(\omega)$ at 
$\omega=0$ and the statistics of the 
sequence of inter-event times $\tau_i$'s (i.e., one plus the lengths
of the cluster of $0$'s between consecutive $1$'s).  For the short-range correlated case, $S(0)$  is finite
and given by \cite{Cox,Benj}:
\begin{equation}\label{eq.spectrum}
S(0)= \frac{\sigma_{\tau}^2}{\langle \tau \rangle^{3}} \left( 1+2\sum_k C_\tau(k) \right).
\end{equation}
For the long-range correlated case, $S(0)\rightarrow\infty$ and Eq.~(\ref{eq.spectrum})
identifies
two different origins: (i) {\it burstiness} measured as the broad tail of the
distribution of inter-event times~$p(\tau)$ (divergent~$\sigma_\tau$); or (ii) long-range
correlations of the sequence of $\tau_i$'s (not summable~$C_{\tau}(k)$).
 In the next section we  show  how these two terms give different
contributions  at different linguistic levels of the hierarchy.

\subsection{Hierarchy of levels}

Building blocks of the hierarchy depicted in Fig.~\ref{fig.1} are binary sequences (organized in levels)  and links between them. 
Levels are established from sets of semantically or syntactically similar conditions $\alpha$'s  (e.g., vowels/consonants, different letters, different
words, different topics)\footnote{Note that our hierarchy of levels  is different from the one used in Ref.~\cite{eckmann1}, which is based on increasingly large adjacent pieces of texts.}.
Each binary sequence~$\textbf{x}$ is obtained by mapping the text using a given $f_\alpha$, and will be denoted by the relevant condition in $\alpha$.
For instance, {\bf prince} denotes the sequence $\textbf{x}$ obtained from the matching condition $\alpha: s_k^{k+7}=$ ``~prince~''. 
A sequence $\textbf{z}$ is linked to $\textbf{x}$ if for all $j$'s such that~$x_j=1$ we
have~$z_{j+r'}=1$, for a fixed constant $r'$. If this condition is fulfilled we say that $\textbf{x}$ is {\it on top of}
$\textbf{z}$ and that $\textbf{x}$ 
belongs to a higher level than $\textbf{z}$. 
By definition, there are no direct links  between sequences at the same level. 
A sequence at a given level is on top of all the sequences in lower levels to which there is a direct path.   
For instance, {\bf prince} is on top of {\bf e} which is on top of {\bf vowel}. 
As will be clear later from our results, the definition of link  can be extended to have a  probabilistic meaning,  suited for
generalizations to high levels (e.g., ``~prince~'' is more probable to appear  while writing about a topic connected to war). 

\subsection{Moving in the hierarchy}\label{ssec.climbing} 
We now show how correlations flow through two linked binary sequences.
 Without loss of generality we denote   ${\bf x}$ a sequence on top of ${\bf z}$ and
$\textbf{y}$ the unique sequence on top of~$\textbf{z}$ such that
 $\textbf{z}=\textbf{x}+\textbf{y}$ (sum and other operations are performed on each symbol: $z_i=x_i+y_i$ for all $i$).   
The spreading of the walker~$Z$ associated with $\textbf{z}$ is given by
\begin{equation}\label{eq.sum}
\sigma^2_Z(t)=\sigma^2_X(t)+\sigma^2_Y(t)+2 C(X(t),Y(t)),
\end{equation}
where $C(A,B)=\langle A B \rangle -\langle A \rangle\langle B \rangle$ is the cross-correlation. 
Using the Cauchy-Schwarz inequality ~$|C(X(t),Y(t))|\le\sigma_X(t) \sigma_Y(t)$ we obtain
\begin{equation}\label{inequality}
\sigma_Z(t) \le \sigma_X(t)+\sigma_Y(t).
\end{equation}
Define  
$\bar{\textbf{x}}$, as the sequence  obtained reverting 
$0\leftrightarrow 1$ on each of its elements $\bar{x}_i=1-x_i$. It is easy to see  that if $\textbf{z}=\textbf{x}+\textbf{y}$ 
then $\bar{\textbf{x}}=\bar{\textbf{z}}+\textbf{y}$. 
Applying the same arguments above, and using that $\sigma_X=\sigma_{\bar{X}}$ for any ${\bf x}$, we obtain  
$
\sigma_X(t) \le \sigma_Z(t)+\sigma_Y(t) $ and similarly $\sigma_Y(t) \le \sigma_Z(t)+\sigma_X(t) $. 
Suppose now that  $\sigma^2_i \simeq t^{\gamma_i}$ with $i \in
\{X,Y,Z\}$. In order to satisfy simultaneously the three inequalities above, at least two out of the three $\gamma_i$ have to be equal to
the largest value $\max_i\{\gamma_i\}$. Next we discuss the implications of this restriction to the flow of
correlations up and down in our hierarchy of levels. 

{\bf Up.}  Suppose that at a given level we have a binary sequence $\textbf{z}$ with long-range correlations~$\gamma_Z>1$. From our
restriction we know that at least one sequence ${\bf x}$ on top of ${\bf z}$, has long-range correlations with
$\gamma_X \ge \gamma_Z$.
This implies, in particular, that if we observe  long-range correlations in the binary sequence associated with a given letter then  we can
argue that its anomaly originates from the anomaly of at least one word where this letter appears, higher  in the hierarchy\footnote{A sequence ${\bf x}$ of a  word containing  the given letter is on top of the sequence ${\bf z}$ of that letter. If 
  ${\bf z}$ is long range correlated (lrc) then  either ${\bf x}$ is lrc or ${\bf y}$ is lrc. Being finite the number of words with a given letter,  we
  can  recursively apply the argument to ${\bf y}$ and identify at least one lrc word.}.

{\bf Down.} Suppose ${\bf x}$ is long-range correlated~$\gamma_X>1$. 
From Eq.~(\ref{eq.sum}) we see that a fine tuning cancellation with cross-correlation must appear in order for their lower-level sequence
${\bf z}$ (down in the hierarchy) to have $\gamma_Z <\gamma_X$. 
From the restriction derived above we know that this is possible only if~$\gamma_X=\gamma_Y$, which is unlikely in the typical case of sequences
$\textbf{z}$  receiving contributions from different sources (e.g., a letter receives contribution from different words). Typically,
  $\textbf{z}$ is composed by $n$ sequences~${\bf x}^{(j)}$, with $\gamma_{X^{(1)}}\neq\gamma_{X^{(2)}}\neq \ldots \neq \gamma_{X^{(n)}}$,
  in which case~$\gamma_Z=\max_j\{\gamma_{X^{(j)}}\}$. Correlations typically flow down in our hierarchy of levels.

{\bf Finite-time effects.}
While the results above  are valid  asymptotically (infinitely long sequences), in the case of any real text we can only have a finite-time estimate~$\hat{\gamma}$ of the correlations $\gamma$. Already from Eq.~(\ref{eq.sum}) we see that the addition of
sequences with different~$\gamma_{X^{(j)}}$, the
  mechanism for moving down in the hierarchy, leads to $\hat{\gamma}_Z < \gamma_Z$ if  $\hat{\gamma}_Z$  is computed at  a time when the asymptotic regime is still not dominating.
This  will play a crucial role in our understanding of long-range correlations in real books. 
In order to give quantitative estimates, we consider the case of ${\bf z}$ being the sum of the most long-range correlated
sequence ${\bf x}$ (the one with $\gamma_X=\max_j\{\gamma_{X^{(j)}}\}$)  and many other independent
non-overlapping\footnote{Sequences $\textbf{x}$ and $\textbf{y}$ are non-overlapping if for
all $i$ for which $x_i=1$ we have $y_i=0$.} sequences whose combined contribution is written as ${\bf y}={\bf \xi}(1-{\bf x})$, with $\xi_i$ an independent
identically distributed binary random variable. This corresponds to the random addition of $1$'s with probability $\langle \xi \rangle$ to
the $0$'s of ${\bf x}$. In this case $\sigma_Z^2$ shows a transition
from normal $\hat{\gamma}_Z=1$ to anomalous $\hat{\gamma}_Z=\gamma_X$ diffusion. The asymptotic regime of {\bf z} starts after a time 
\begin{equation}\label{eq.tt}
t_T \ge  \left(\frac{\langle \xi \rangle}{1-\langle \xi \rangle} \frac{1}{g \langle x \rangle}\right)^{1/(\gamma_X-1)}, 
\end{equation}
where $0< g\le 1$ and $\gamma_X>1$ are obtained from   $\sigma^2_X $ which  asymptotically goes as  $g\langle x  \rangle \langle
  1-x\rangle  t^{\gamma_X}$. Note that the power-law sets at $t=1$ only if $g=1$.
A similar relation is obtained moving
up in the hierarchy, in which case a sequence ${\bf x}$ in a higher level is built by 
random subtracting $1$'s from the lower-level sequence ${\bf z}$ as ${\bf x}= {\bf \xi z}$ (see SI-Sec.~III-A for all calculations).

{\bf Burstiness.}
In contrast to correlations, burstiness due to the tails of the inter-event time distribution~$p(\tau)$ is not always preserved when
moving up and down in the hierarchy of levels. Consider first going down by adding sequences with different tails of $p(\tau)$. The tail of the combined
sequence will be constrained to the shortest tail of the individual sequences. In the random addition example, ${\bf z}={\bf
  x}+{\bf \xi}(1-{\bf x})$ with ${\bf x}$ having a broad tail in~$p(\tau)$,  the large $\tau$ asymptotic of ${\bf z}$ has short-tails because  the cluster of zeros
in ${\bf x}$ is cut randomly by $\xi$~\cite{allegrini2}.  Going up in the hierarchy, we take a sequence on top
of a given bursty binary sequence, e.g., using the random subtraction ${\bf x}={\bf \xi} {\bf z}$ mentioned above.
The probability of finding a large inter-event time $\tau$ in ${\bf z}$ is enhanced  by the
number of times the random deletion merges two or more clusters of $0$'s in ${\bf x}$, and diminished by the
number of times the deletion destroys a previously existent inter-event time $\tau$. Even accounting for the change in $\langle \tau \rangle$, this moves cannot
lead to a short-ranged~$p(\tau)$ for ${\bf x}$ if $p(\tau)$ of ${\bf z}$ has a long tail (see SI-Sec.~III-B). Altogether, we expect
burstiness to be preserved moving up, and destroyed moving down in the hierarchy of levels.
 
{\bf Summary.}
From Eq.~(\ref{eq.spectrum})  the origin of long-range correlations~$\gamma>1$  
can be traced back to two different sources:  the  tail of 
$p(\tau)$ 
(burstiness) and the tail of $C_\tau(k)$.  
The computations above  reveal  their different role at different levels in the hierarchy: 
$\gamma$ is preserved moving down, but there is a transfer of {\it information} from
$p(\tau)$ to $C_\tau(k)$.
This is better understood by considering the following simplified set-up: suppose at a given level we observe a
sequence~${\bf x}$ coming from a renewal process with broad tails in the inter-event times 
\begin{equation}\label{eq.renewal}
p(\tau) \sim \tau^{-\mu} \text{ and } C_\tau(k)=\delta(k),
\end{equation}
with $2<\mu<3$ leading to~$\gamma_X=4-\mu$~\cite{allegrini}. Let us now consider what is observed in {\bf z}, at
a level below, obtained by adding to~${\bf x}$ other independent sequences. The long~$\tau$'s (a long sequence of 0's) in Eq.~(\ref{eq.renewal})
will be split in two long sequences introducing at the same time a cut-off $\tau_{c}$ in~$p(\tau)$ and non-trivial correlations
$C_\tau(k)\neq0$ for large~$k$.   
In this case, asymptotically the long-range correlations ($\gamma_Z=\max\{\gamma_X,\gamma_Y\}>1$) is solely due to
$C_\tau(k)\neq0$. Burstiness affects only~$\hat{\gamma}$ estimated for times~$t<\tau_c$. 
A similar picture is expected in the generic case of a starting sequence~${\bf x}$ with 
broad tails in both $p(\tau)$ and $C_\tau(k)$.


\begin{figure}[!bt]
\includegraphics[width=1\columnwidth]{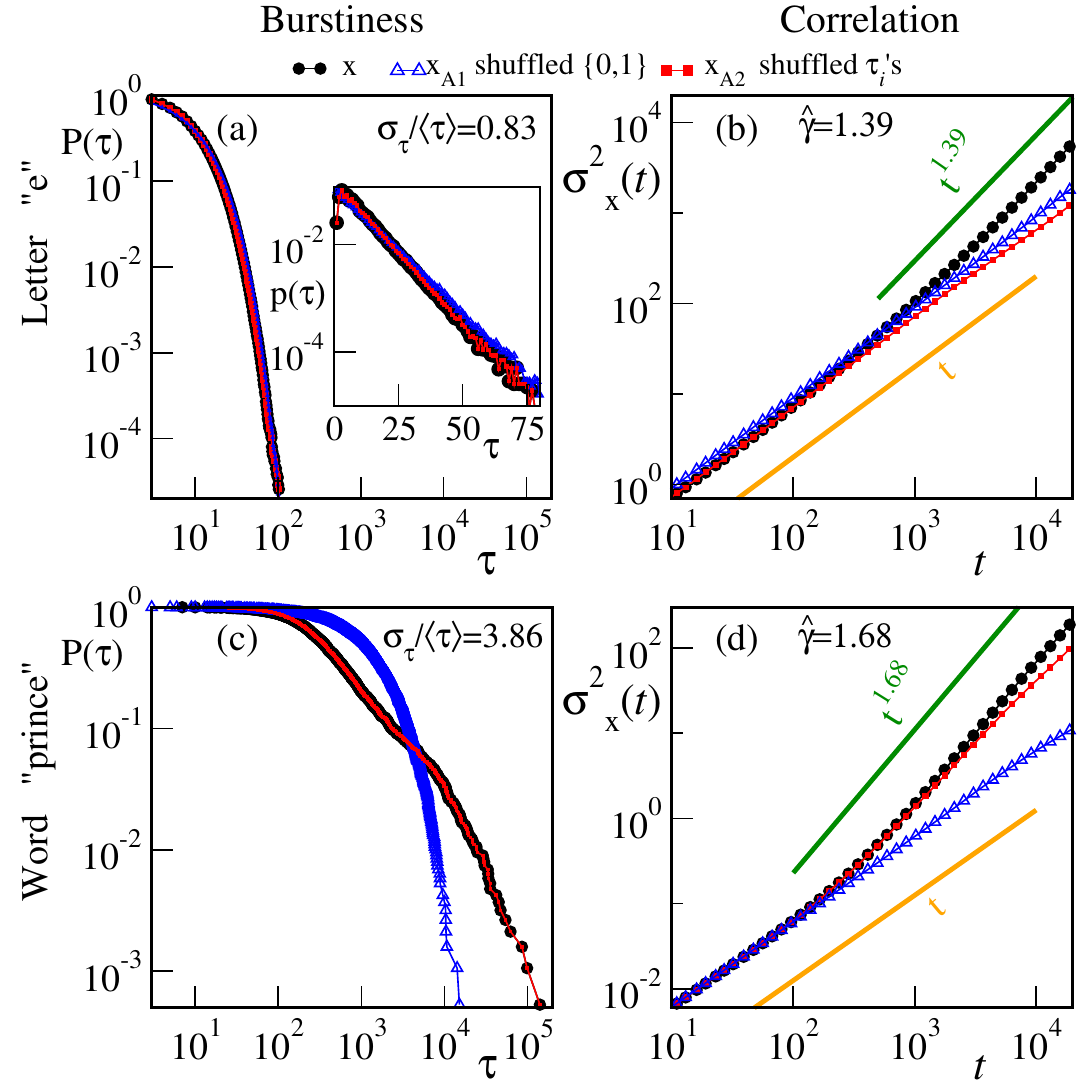}
\caption{ Burstiness and long-range correlation on different linguistic levels. The binary sequences of the letter ``e'' (a,b) and of the
  word `` prince '' (c,d) in the book ``War and Peace'' are shown. (a,c) The cumulative inter-event time distribution $P(\tau)\equiv\int_0^\tau p(t')dt'$. (b,d) Transport~$\sigma^2_X(t)$ defined in 
  Eq.~(\ref{eq.mu}). The numerical results show: (a) exponential decay of $P(\tau)$  with $\sm=0.83$
 Inset: $p(\tau)$ in log-linear scales; (b) $\hat{\gamma}=1.39\pm0.05$; (c) non-exponential
  decay of $P(\tau)$ with $\sm=3.86$; and (d)~$\hat{\gamma}=1.68\pm0.05$.  All panels show results for the the original and
  $A_1,A_2$-shuffled sequences, see legend. 
}\label{fig.2}
\end{figure}

\begin{figure}[!ht]
\includegraphics[width=\columnwidth]{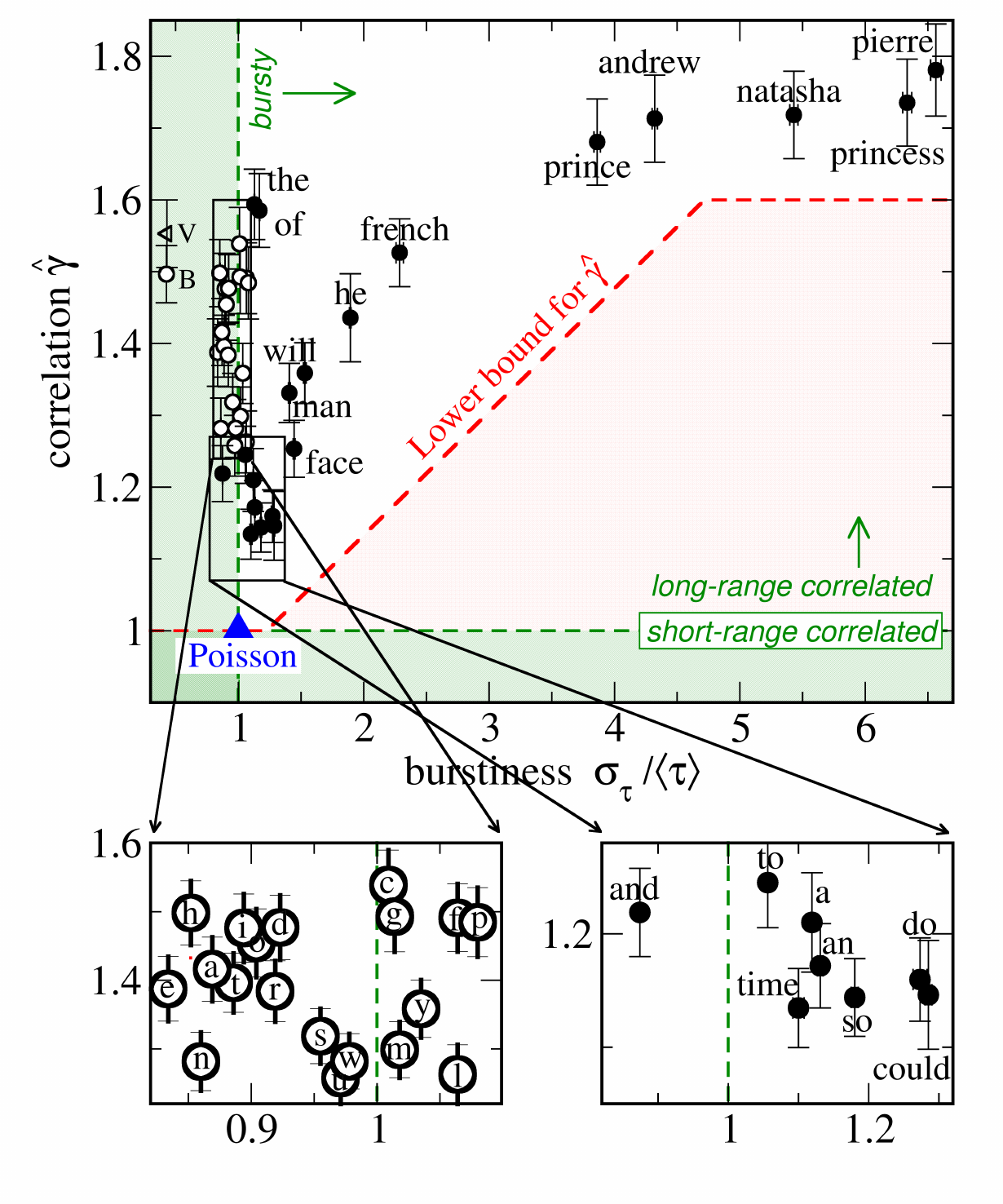}
\caption{Burstiness-correlation diagram for sequences at different levels. $\sm$ is an indicator of the
  burstiness of the distribution~$p(\tau)$. $\hat{\gamma}$ is a finite time estimator of the global indicator of long-range
  correlation~$\gamma$. A Poisson process has $(\sm,\gamma)=(1,1)$.
The twenty most frequent symbols (white circles) and twenty frequent words (black circles) of wrnpc are shown (see 
  SI-Tables for all books). $V$ indicates the case of {\bf vowels} and $B$ of {\bf blank space}. The red dashed-line is a
  lower-bound estimate of $\hat{\gamma}$ due to burstiness (see SI-Sec. VI). This diagram is a generalization for long-range correlated
  sequences of 
  the diagrams in Ref.~\cite{goh}.} \label{fig.3}  
\end{figure}

\section{Data analysis of literary texts}\label{sec.observations}

Equipped with previous section's theoretical framework, here we interpret observations in
real texts. We use ten English versions of international novels (see SI-Sec.~IV for the list and for the pre-processing applied to the texts). For each
book $41$ binary sequences were analyzed separately: 
vowel/consonants, $20$ at the letter level (blank space and the $19$ most frequent 
letters), and $20$ at the word  level ($6$ most frequent words, $7$ most frequent nouns, and $7$ words with frequency matched to the
frequency of the  nouns). The finite-time estimator of the long-range correlations~$\hat{\gamma}$ was computed fitting
Eq.~(\ref{eq.mu}) in a broad range  of large $t \in [t_{s'},t_s]$ (time lag of correlations) up to $t_s=1\%$ of the book size. This range was obtained using a conservative procedure designed to
robustly distinguish between short and long-range correlations (see SI-Sec.~V). 
We illustrate the results in our longest novel, ``War and Peace''  by L. Tolstoy (wrnpc, in short, see
SI-Tables for the results in all books).

\subsection{Data analysis of correlations and burstiness} One of the main goals of our measurements is to distinguish, at  different hierarchy levels, between
the two possible sources of long-range correlations in Eq.~(\ref{eq.spectrum}) -- burstiness corresponding to $p(\tau)$ with diverging
$\sigma_\tau$ or diverging $\sum C_\tau(k)$. To this end we compare the results with two null-model binary sequences ${\bf x}_{A1}, {\bf
  x}_{A2}$ obtained by applying to ${\bf x}$ the following procedures: 
\begin{itemize}
\item [A1:] shuffle the sequence of~$\{0,1\}$'s. Destroys all correlations. 
\item [A2:] shuffle the sequence of inter-event times~$\tau_i$'s. Destroys correlations due to $C_\tau(k)$ but preserves those due to~$p(\tau)$.
\end{itemize}
Starting from the lowest level of the hierarchy depicted in Fig.~\ref{fig.1}, we obtain $\hat{\gamma}=1.55 \pm 0.05$ for the sequence
of vowels in wrnpc 
and~$\hat{\gamma}$ between~$1.18$ and~$1.61$ in the other~$9$ books (see SI-Fig.~S1).
The values for ${\bf x_{A1}}$ and ${\bf x_{A2}}$ were compatible (two error bars) with the expected value~$\gamma=1.0$ in all
books. 
Figures~\ref{fig.2}ab show the computations for the
case of the letter~``e'': while 
$p(\tau)$ decays exponentially in all cases (Fig.~\ref{fig.2}a), long-range
correlations are present in the original sequence ${\bf e}$ but absent from the A2 shuffled version of ${\bf e}$ (Fig.~\ref{fig.2}b). This
means that burstiness is  
absent from ${\bf e}$ and does not contribute to its long-range correlations. 
In contrast,  for the word ``~prince~'' Fig.~\ref{fig.2}c shows a non-exponential~$p(\tau)$ and Fig.~\ref{fig.2}d shows that the original
sequence ${\bf prince}$ and the~$A2$  shuffled sequence show similar long-range correlations (black 
and red curves, respectively).  This means that the origin of the
long-range correlations of {\bf prince} are mainly due to burstiness -- tails of~$p(\tau)$ -- and not to
correlations in the sequence of~$\tau_i$'s -- $C_\tau(k)$. 

In Fig.~\ref{fig.3} we plot for different sequences the summary quantities $\hat{\gamma}$ and $\sm$ -- a measure of the burstiness proportional to
the relative width of~$p(\tau)$~\cite{ortuno,goh}. A Poisson process has $\gamma=\sm=1$. All {\it letters} have $\sm\approx1$, but clear
long-range correlations $\hat{\gamma}>1.1$ (left box magnified in Fig.~\ref{fig.3}). 
This means that correlations come from~$C_\tau(k)$ and not from~$p(\tau)$, as shown in Fig.~\ref{fig.2}(a,b) for the letter ``e''. 
The situation is more interesting in the higher-level case of {\it words}. 
The most frequent words and the words selected to match the nouns mostly show $\sm\approx1$ so that the same conclusions we drew
about letters apply to these words. In contrast to this group of function words are the most frequent {\it nouns}
that have large~$\sm$~\cite{allegrini,herrera,ortuno,altmann} and large~$\hat{\gamma}$, appearing as outliers at the upper right corner of Fig.~\ref{fig.3}. The case of ``~prince~'' shown in
Fig.~\ref{fig.2}(c,d) is representative of these words, for which burstiness contributes to the long-range correlations. 
In order to confirm the generality of Fig.~\ref{fig.3} in the $10$ books of our database, we performed a pairwise comparison of $\hat{\gamma}$
and $\sigma_\tau/\langle \tau \rangle$ between the $7$ nouns and their frequency matched words.  Overall, 
the nouns had a larger $\hat{\gamma}$ in $56$ and a larger $\sigma_\tau /\langle \tau \rangle$ in $55$ out of the $70$ cases (P-value~$<10^{-6}$,
assuming equal probability). In every single book at least $4$ out of $7$ comparisons show larger values of $\hat{\gamma}$ and $\sm$ for the
nouns. 

 We now explain a striking feature of  the data shown in Fig.~\ref{fig.3}: the absence of sequences with low $\hat{\gamma}$ and high
  $\sm$ (lower-right corner). This
is an evidence of correlation between these two indicators and motivates us to estimate a $\sm$-dependent
  lower bound for  $\hat{\gamma}$, as shown in Fig.~\ref{fig.3}. Note that   
  high values of burstiness  are responsible for  long-range correlations estimate $\hat{\gamma} >1$, as discussed  after Eq.~(\ref{eq.renewal}).
   For   instance, the  slow decay of~$p(\tau)$ for intermediate~$\tau$ in {\bf prince} (Fig.~\ref{fig.2}c) leads to $\sm \gg 1$ and an estimate
  $\hat{\gamma}>1$  at intermediate times. 
Burstiness contribution  to $\hat{\gamma}$  (which gets also contributions from long-range correlations in the $\tau_i$'s) is measured by
  $\hat{\gamma}_{A2}$, which is usually a lower bound for the total long-range correlations:  $\hat{\gamma}\ge \hat{\gamma}_{A2}$. 
More quantitatively, consider an $A2$-shuffled sequence with power-law  $p(\tau)$ --  as in  Eq.~(\ref{eq.renewal}) -- with
  an exponential cut-off for $\tau>\tau_c$.  By increasing $\tau_c$  we have that $\sm$ monotonously increases [it can be
 computed directly  from $p(\tau)$]. In terms of $\hat{\gamma}_{A2}$, if the fitting interval $t \in [t_{s'},t_s]$ used to compute the finite time
 $\hat{\gamma}_{A2}$ is all below $\tau_c$ (i.e. $ t_s < \tau_c$) we have $\hat{\gamma}_{A2}= 4-\mu > 1$ (see Eq.~(\ref{eq.renewal})) while
 if the fitting interval is all beyond the cutoff (i.e. $\tau_c < t_{s'}$ ) we have  $\hat{\gamma}_{A2}=1$. Interpolating linearly between these two values and using
 $\mu=2.4$ we obtain the lower bound for $\hat{\gamma}$ in Fig.~\ref{fig.3}.
It strongly restricts the range of possible $(\sm,\hat{\gamma})$ in agreement with the observations and also with  
$\hat{\gamma}$ obtained for the $A2$-shuffled sequences (see SI-Sec. VI for further details).

\subsection{Data analysis of finite-time effects}

\begin{figure}[!ht]
\includegraphics[width=\columnwidth]{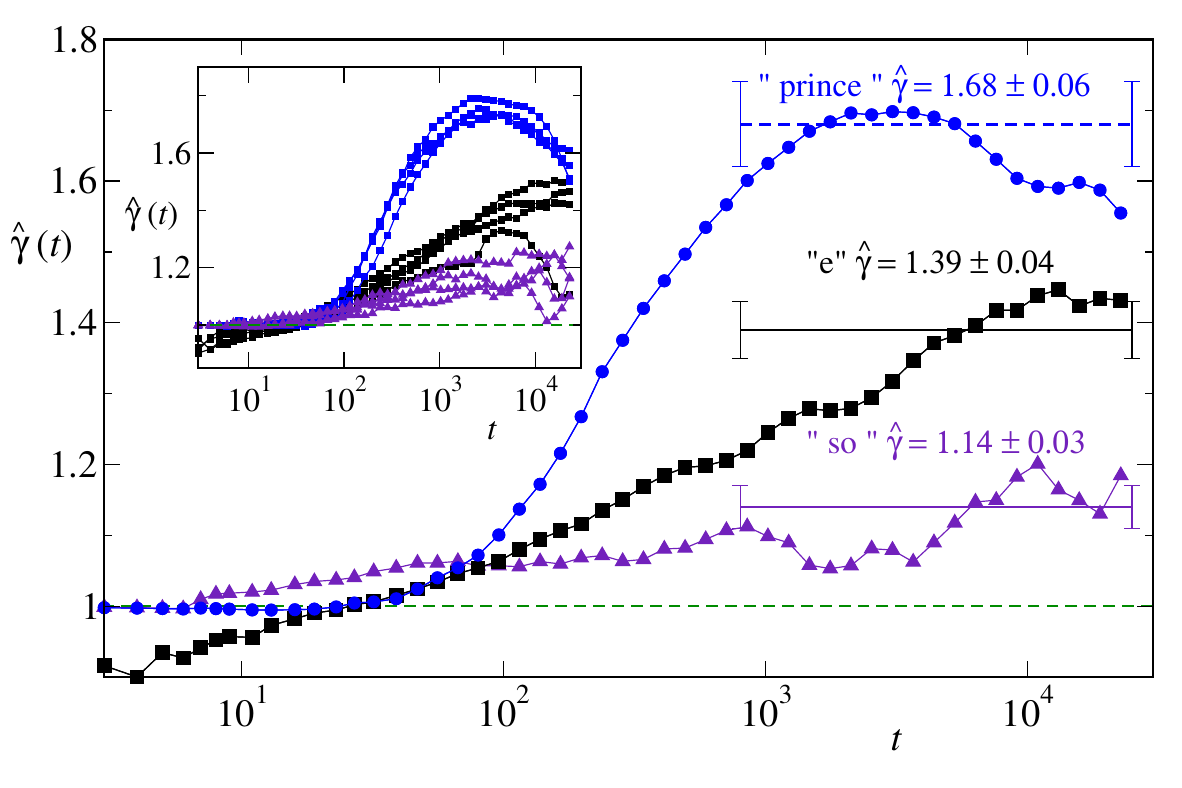}
\caption{ Transition from normal to anomalous behavior. The time dependent exponent is computed as  $\hat{\gamma}(t) \equiv \Delta \log \sigma_X^2(t)/\Delta \log t$ (local derivative of the transport
curve in Fig.~\ref{fig.2}bd). Results for three sequences in wrnpc are shown (from top to bottom): the noun ``~prince~'', the most frequent
letter ``e'', and the word ``~so~'' (same frequency of`` prince ''). The horizontal lines indicate the $\hat{\gamma}$, the error bars, and the fitting range. Inset (from top to bottom): the $4$ other nouns appearing as outliers in Fig.~\ref{fig.3}, the $4$ most frequent
  letters after ``e'', and the $4$ words matching the frequency of the outlier-nouns. 
}\label{fig.4}
\end{figure}

The pre-asymptotic normal diffusion  -- anticipated in Sec. {\bf Finite-time effects} -- is clearly 
seen in Fig.~\ref{fig.4}. Our theoretical model explains also other specific observations:

\noindent 1. Key-words reach higher values of~$\hat{\gamma}$ than letters ($\hat{\gamma}_{\text{e}} < \hat{\gamma}_{\text{prince}}$). This
observation contradicts our expectation for asymptotic long times: {\bf prince} is on top of {\bf e} and the reasoning after
Eq.~(\ref{eq.sum}) implies $\gamma_{\text{e}} \ge \gamma_{\text{prince}} $. This seeming contradiction is solved by our estimate~(\ref{eq.tt}) of the  transition time $t_T$ needed for the finite-time estimate $\hat{\gamma}$ to reach the asymptotic $\gamma$. This is done imagining a 
surrogate sequence with the same frequency of ``e'' composed by {\bf prince}  and randomly added $1$'s. Using the fitting values of~$g,\gamma$ for {\bf prince} in Eq.~(\ref{eq.tt}) we obtain
$t_T \ge 6\;10^5$, which is larger than the maximum time $t_s$ used to obtain~$\hat{\gamma}$. Conversely, for a
sequence with the same frequency of ``~prince~'' built as a random sequence on top of {\bf e} we obtain $t_T \ge 7\;10^8$. These
calculations not only explain {$\hat{\gamma}_{\text{e}} < \hat{\gamma}_{\text{prince}}$, they show} 
 that {\bf prince} is a particularly meaningful (not random) sequence on top of {\bf e}, and that {\bf e} is necessarily composed by other sequences with~$1<\gamma<\hat{\gamma}_{\text{prince}}$ that dominate for shorter times. More generally, the {\it observation} of
long-range correlations at low levels is due to widespread correlations on higher levels.

\noindent 2. The sharper transition for keywords. The addition of many sequences with $\gamma>1$ explains
the slow increase in $\hat{\gamma}(t)$ for letters because sequences with increasingly larger $\gamma$ dominate for increasingly longer
times. The same reasoning explains the positive correlation between $\hat{\gamma}_e$  and the length of the book (Pearson 
Correlation~$r=0.44$, similar results for other letters). 
The sequence ${\bf so}$ also shows slow transition and small~$\hat{\gamma}$, consistent with the interpretation that it is connected to many
topics on upper levels. In contrast, the sharp transition for {\bf prince} indicates the existence of fewer independent contributions on higher levels, consistent with the observation of the onset of burstiness $\sm>1$. Altogether, this strongly supports our model of hierarchy of levels  with keywords (but not function words) strongly connected to specific topics which are the actual correlation
carriers. The sharp transition for the keywords appears systematically roughly at 
the scale of a paragraph ($10^2-10^3$ symbols), 
in agreement with similar observation in Refs.~\cite{montemurro,melnyk2005,eckmann1,doxas}.

\subsection{Data analysis of shuffled texts}
Additional insights   on  long-range correlations are obtained by investigating  whether they are robust  under different manipulations of the text~\cite{ebeling3,eckmann1}. Here we focus on two non-trivial shuffling methods (see SI-Sec.~VII for simpler cases for which our theory leads to analytic results). Consider generating new same-length texts by applying to the original texts the following procedures

\begin{itemize}
\item [M1] Keep the position of all blank spaces fixed and place each word-token randomly in a gap of the size of the word.
\item [M2] Recode each word-type by an equal length random sequence of letters and replace consistently all its tokens.
\end{itemize}
Note that M1 preserve structures (e.g., words and letter frequencies) destroyed
by M2. In terms of our hierarchy, M1 destroys the links to levels above word level while M2 shuffles the links from word- to letter-levels.
Since according to our picture correlations originate from high level structures, we predict that M1 destroys and M2 preserves long-range
correlations. 
Indeed simulations unequivocally show that long-range correlations present in the original texts (average~$\hat{\gamma}$ of letters in wrnpc $1.40 \pm 0.09$ and in
all books $1.26\pm 0.11$) are mostly destroyed by M1 ($1.10 \pm 0.08$ and $1.07\pm0.08$) and preserved by M2 ($1.33\pm0.08$ and
$1.20\pm0.09$ (see SI-Tables for all data). 
At this point it is interesting to draw a connection to   the {\it principle of the arbitrariness of the sign}, according to which the association between a given sign (e.g., a word) and the referent (e.g., the object in the real world) is arbitrary~\cite{saussaure}. As confirmed by the M2 shuffling, the long-range correlations of literary texts are invariant under this principle because they are connected to the semantic of the text. Our theory is consistent with this principle.

\section{Discussion} 
From an information theory viewpoint, long-range correlations in a symbolic sequence have two different and concurrent
sources:  the broad distribution of the distances between successive occurrences of the same symbol (burstiness) and the correlations of
these distances. We found  that the contribution of these two sources is very different for observables of a literary text at
different linguist levels. In particular, our theoretical framework provides a robust mechanism explaining our extensive observations that
on relevant semantic levels the text is high-dimensional and bursty while on lower levels successive projections destroy burstiness while preserving
the long-range correlations of the encoded text via a flow of information from burstiness to correlations.  

The mechanism explaining how correlations cascade from high- to low-levels
is generic and extends to levels higher than word-level in the hierarchy in Fig.~\ref{fig.1}.  
The construction of such levels could be based, e.g., on techniques devised to extract information on a ``concept space''
  \cite{eckmann1,doxas,montemurro}.  While long-range correlations have been observed at the concept level~\cite{eckmann1}, further studies are required to connect to observations made at lower levels and to distinguish between the two sources of correlations. 
Our results showing that correlation is preserved after random additions/subtractions of $1$'s help this connection because they show that words can be linked to concepts even  if they are not used every single time the concept appears (a high probability suffices).
For instance, in Ref.~\cite{eckmann1} a topic can be associated to an axis of the concept space and be linked to the words used to
build it. In this case, when the text is referring to a topic there is a higher probability of using the words linked to it and therefore
our results show that correlations will flow from the topic to the word level.
In further higher levels, it is insightful to consider as a limit picture the renewal case --
Eq.~(\ref{eq.renewal}) -- for which long-range correlations originate only due to burstiness. This {\it limit case} is the
simplest toy model compatible with our results. Our theory predicts that correlations take form of a bursty sequence of events once we approach the semantically relevant topics of the text. Our observations show that some highly topical words already show long-range correlations mostly due to burstiness, as expected by observing that topical words are connected to less concepts than function words~\cite{altmann}.
This renewal limit case is the desired outcome of successful analysis of anomalous diffusion in dynamical systems and has been speculated to appear in various fields~\cite{allegrini,allegrini2}.  Using this limit case as a guideline we can  think of an algorithm able to 
automatically detect the relevant structures in the hierarchy by pushing recursively the long-range correlations
into a renewal sequence.

Next we discuss how our results improve previous analyses and open new possibilities of applications. 
Previous methods either worked below the letter level~\cite{grassberger,schenkel,kokol,kanter} or combined the correlations of different
letters in such a way that asymptotically the most long-range correlated sequence dominates~\cite{voss,ebeling1,ebeling3}. Only through our
results it is possible to understand that indeed a single asymptotic exponent $\gamma$ should be expected in all these cases. However, and
more importantly, $\gamma$ is usually beyond observational range and an interesting range of finite-time~$\hat{\gamma}$ is 
obtained depending on the observable or encoding. On the letter level, our analysis  (Figs.~\ref{fig.2} and~\ref{fig.3}) revealed that all
of them are long-range  correlated with no burstiness (exponentially distributed inter-event times). This lack of burstiness can be wrongly
interpreted as an indication that letters~\cite{goh} and most parts of speech~\cite{badalamenti}  
are well described by a Poisson processes. Our results explain that the non-Poissonian (and thus information rich) character of the text is
preserved in the form of long-range correlations~($\gamma>1$), which is observed also for all frequent words (even in the most frequent word
``~the~''). These observations 
violate not only the strict assumption of a Poisson process, they are incompatible with any finite-state Markov chain model.
These models are the basis for numerous applications of automatic semantic information extraction, such as keywords extraction, authorship
attribution, plagiarism detection, and automatic summarization~\cite{manni,ober,usatenko,stamata}. All these applications can potentially benefit
from our deeper understanding of the mechanisms leading to long-range correlations in texts. 

Apart from these applications, more fundamental extensions of our results should: (i) consider the mutual information and similar
entropy-related quantities, which have been widely used to quantify long-range correlations~\cite{li,ebeling3} 
(see~\cite{HerGro95} for a comparison to correlations); (ii) go beyond the simplest case of the two point
autocorrelation function and consider multi-point correlations or higher order entropies~\cite{ebeling3}, which are necessary for
the complete characterization of the correlations of a sequence; and (iii) consider the effect of
 non-stationarity on higher levels, which could cascade to lower levels and affect correlations properties.
 Finally, we believe that our approach may help to understand long-range correlations in
any complex system for which an hierarchy of levels can be identified, such as human activities~\cite{rybski} and DNA
sequences~\cite{li,peng,voss,ScEbHe96}.

\section*{Acknowledgments}
We thank B. Lindner for insightful suggestions and S. Graffi for the careful reading of the manuscript. G.C.  acknowledges partial support by the FIRB-project RBFR08UH60 (MIUR, Italy). M. D. E. acknowledges partial support by the PRIN project 2008Y4W3CY (MIUR, Italy).



\newpage

\setcounter{figure}{0}
\setcounter{section}{0}
\begin{widetext}
\begin{center}
{{\sc \huge Supporting Information  }}
\end{center}
\end{widetext}

\section{Average procedure in binary sequences}
Given an ergodic and stationary stochastic process, correlation functions are defined as 
\begin{equation}\label{eq.corr}
\textrm{Corr}(j,t):=E\left( x_j x_{j+t} \right) -\textrm{E} (x_j )\textrm{E}( x_{j+t} ).
\end{equation}
where $E(\cdot)$ denotes an average over different realizations~${\bf x}$ of the process. Stationarity guarantees that $\textrm{Corr}(j,t)$ depends on the time lag $t$ only.
In practice, one typically has no access to different realizations of the process but only to a single finite sequence. In our case, any binary
sequence \textbf{x}  is  obtained from a single text of length  $N$ through a given mapping. In such cases it is possible to use the assumption of
ergodicity to approximate the  correlation function~(\ref{eq.corr}) by
\begin{equation*}
C_x(t):= \langle x_jx_{j+t} \rangle -\langle x_j \rangle\langle x_{j+t} \rangle,
\end{equation*}
where $\langle \cdot \rangle$ means averaging, for each fixed $t$, over all pairs $x_j$ and $x_{j+t}$
for $j=1,2,..., (N-t)$ as
\begin{equation*}
\langle \cdot \rangle \equiv \frac{1}{N-t} \sum_{j=1}^{N-t} \cdot.
\end{equation*}

\section{Mapping  Examples}
 Consider the sentence ``This paper is a paper of mine''. By choosing  the condition $\alpha$ to
be  \textit{the $k$-th symbol is a vowel } the projection $f_{\alpha}$ maps the sentence into the sequence $\{
00100010100100100101001000101\}$. If $\alpha$ is \textit{the k-th symbol is equal to `e'} than 
we get: $\{0000000010000000100001000000001\}$. Generally, we can treat any n-gram of letters in the same way, as for example by choosing the
condition $\alpha$ to be \textit{ the $2$-gram starting at the k-th symbol is equal to `er'}, that projects using a sliding window the sentence to:$\{00000000100000000001000000000\}$. 
Words are encoded using their corresponding n-gram, for example $\alpha$ could be \textit{ the 7-gram starting at the k-th
  symbol is equal to ` paper ' (blank spaces included)} that gives: $\{0000100000000000010000000\}$. 
It is possible to
generalize these procedures to more \textit {semantic conditions} $\alpha$ that  associate $1$ to either all or part of the 
symbols that appears in a sentence that is attached to a  specified topic. These topics can be quantitatively constructed from the frequency
of words using methods such as latent semantic analysis~\cite{landauer} or the procedures to determine the so-called {\em concept space}~\cite{eckmann1}.

\section{Simple operations on binary sequence and their effects on long-range correlations and burstiness} \label{sec.operations}
We describe two simple procedures to construct two binary sequences {\bf x} and {\bf z} such that {\bf x} is {\it on top of} {\bf z}.
These procedures will be based either on the ``addition" of $1$'s to {\bf x} or on the ``subtraction" of $1$'s of {\bf z}. In the
simplest cases of {\it random} addition and subtraction, we explicitly compute how long-range  correlations flow from  {\bf x} to {\bf z}  (corresponding to a flow from upper to lower 
levels of the hierarchy) and how burstiness is preserved when extracting {\bf x} from {z} ( moving from lower to upper levels in the hierarchy).  

Recall that a sequence {\bf x} is {\em on top of} {\bf z}  if for all $j$  such that $x_j =1$ we have $z_{j+r} = 1$, for a fixed constant
$r$. Without loss of generality in the following calculations we fix for simplicity $r=0$.
We now define simple operations that map two binary sequences into a third binary sequence:

\begin{itemize}

\item Given two generic binary sequences  {\bf z} and {\bf $\xi$} we define their multiplication {\bf y}={\bf $\xi$} {\bf z} as 
  $ y_i= \xi_i z_i, \, \forall i$.  
By construction {\bf y} is on top of {\bf z}.

\item Given two non-overlapping  sequences  {\bf x} and  {\bf y } we define their sum {\bf z} =  {\bf x}+{\bf y}  as 
$z_i=x_i+y_i, \, \forall i$. 
By construction {\bf x } and {\bf y} are on top of {\bf z}. We say  that sequences $\textbf{x}$ and $\textbf{y}$ are non-overlapping if for
all $i$ for which $x_i=1$ we have $y_i=0$.

\end{itemize}

In general,  two independent binary sequences  {\bf x} and  {\bf $\xi$}  will overlap. A sequence $\textbf{y}$ which is non-overlapping with $\textbf{x}$ can be constructed  from {\bf $\xi$} as {\bf y}= {\bf$\xi$}({\bf 1}-{\bf x}), where {\bf 1} denotes the trivial sequence with all $1$'s.
In this case, we say that   {\bf z} = {\bf x}+{\bf y},   with {\bf y} ={\bf$\xi$}({\bf 1}-{\bf x})  is  a sequence lower than {\bf x} in the hierarchy  that is constructed by  a {\it random addition} (of 1's)  to {\bf x}. 
Similarly, if {\bf $\zeta$} is independent of {\bf z}, the sequence {\bf $\zeta$}{\bf z} is a {\it random subtraction} (of 1's)  of {\bf z}  

\subsection{Transition time from normal to anomalous diffusion}
Consider a sequence {\bf z} constructed as a random addition of $1$'s to a given long-range correlated sequence ${\bf x}$: ${\bf z}={\bf
  x}+{\bf y}$,   with {\bf y} ={\bf$\xi$}({\bf 1}-{\bf x}) and  $\xi$  a sequence of {\it i.i.d.} binary random variables.
The associated random walker $Z$ spreads anomalously with the same exponent of $X$. 
This asymptotic regime is masked at short times by a pre-asymptotic normal behavior.
Here we first compute explicitly the spreading of $Z$ in terms of that of $X$ and $Y$ and then we compute a bound for the transition time
$t_T$ to the asymptotic anomalous diffusion of $Z$. 

As written in Eq.~(5)  of the main text we have
\begin{equation}\label{eq.sum}
\sigma_Z^2(t)=\sigma_X^2(t)+\sigma_Y^2(t) + 2C(X(t),Y(t)).
\end{equation}
For our particular case we obtain 
\begin{equation}\label{eq.Y}
\langle Y(t) \rangle = \langle \xi \rangle t \left(1-\langle x\rangle\right).
\end{equation}
and 

\begin{widetext}
\begin{eqnarray}
\langle Y(t)^2\rangle&=& \left< \sum_{i,j=1}^t (\bx_i\xi_i)( \bx_j \xi_j) \right> \nonumber \\
&=&\left< \sum_{i=1}^t (\bx_i^2\xi_i^2)\right> + \left< \sum_{i,j=1,i\neq j}^t (\bx_i\xi_i)( \bx_j \xi_j) \right>\nonumber \\ 
&=& \sum_{i=1}^t \left <\bx_i^2\right> \left <\xi^2\right> + \sum_{i,j=1, i\neq j}^t \left< \bx_i  \bx_j  \right> \langle \xi\rangle^2 \nonumber \\ 
&=& \sum_{i=1}^t \left <\bx_i^2\right> \left <\xi^2\right> -\sum_{i=1}^t \left <\bx_i^2\right> \left <\xi\right>^2 + \quad \sum_{i=1}^t \left <\bx_i^2\right> \left <\xi\right>^2 + \sum_{i,j=1,i\neq j}^t \left< \bx_i  \bx_j  \right> \langle \xi \rangle^2 \nonumber \\ 
&=&\langle \xi \rangle^2\langle X(t)^2\rangle+\sigma^2(\xi)  \sum_{i=1}^t \langle \bx_i^2\rangle \label{squared-app}.
\end{eqnarray}
From Eqs.~(\ref{eq.Y}) and~(\ref{squared-app}) -- and noting that $  \sum_{i=1}^t \langle \bx_i^2 \rangle =   \sum_{i=1}^t \langle \bx_i \rangle = t (1-
\langle x \rangle) $ and  $\sigma_{\bar{X}}^2(t)=\sigma_X^2(t)$ -- we obtain 
\begin{equation}\label{eq.sigma2}
\sigma_Y^2(t) \equiv \langle Y^2(t) \rangle - \langle Y(t) \rangle^2 = \langle \xi\rangle^2 \,\sigma^2_X(t)+t\,\sigma^2_\xi (1-\langle x\rangle)
\end{equation}
The correlation term in Eq.~(\ref{eq.sum}) can also be obtained through direct calculations:
\begin{eqnarray}
C(X(t),Y(t))&=&\left< X(t)Y(t)\right> -\left< X(t)\right> \left< Y(t)\right>\nonumber\\
&=&\left< \sum_{i,j=1}^t x_i(1-x_j)\xi_j \right> - \left< X\right>  \left< \sum_{j=1}^t  (1-x_j)\xi_j\right> \nonumber\\
&=&\left< X(t)\right>  \left<\xi\right> t - \left< X^2(t)\right>  \left<\xi\right>  \nonumber  -  \left< X(t)\right>\Big[  \left<\xi\right> t  -\left< X(t)\right> \left<\xi\right> \Big]\\
&=&-\left< \xi \right> \sigma^2_X(t).\label{eq.correl}
\end{eqnarray}
\end{widetext}
Finally, inserting Eqs.~(\ref{eq.sigma2}) and (\ref{eq.correl}) into Eq.~(\ref{eq.sum}) we have
\begin{eqnarray}
\sigma_Z^2(t)&=&\sigma_X^2(t)+\sigma_Y^2(t) + 2C(X(t),Y(t))\nonumber\\
&=& \sigma_X^2(t)+\langle\xi\rangle^2\sigma_X^2(t)+t\sigma_\xi^2 (1-\langle x \rangle)-2\langle\xi\rangle \sigma_X^2(t)\nonumber\\
&=& t  \sigma_\xi^2 (1-\langle x\rangle) + \sigma_X^2(t)(1-\langle \xi\rangle)^2\nonumber\\
&=& \langle\xi\rangle(1-\langle \xi\rangle)(1-\langle x\rangle) t \,+\, (1-\langle\xi\rangle)^2\sigma_X^2(t)\label{eq.sumsimple}
\end{eqnarray}
As $X$   superdiffuses   so it will $Z$ and they both have the same asymptotic behavior. On the other hand   the asymptotic regime is masked at short times by a pre-asymptotic normal behavior, given by the 
  linear term in $t$. We stress that, even if the non-overlapping condition for {\bf y} forces both $\sigma^2_Y(t)$  and $C(X(t),Y(t))$ to have the same asymptotic behavior of $\sigma^2_X(t)$, their cumulative contributions does not cancel out  unless we trivially have $\langle \xi \rangle =1$. 
  \newline

We  now give a bound on the transition time~$t_T$ to the asymptotic anomalous diffusion of Eq.~(\ref{eq.sumsimple}).
 Without loss of generality consider the case in which even the asymptotic anomalous behavior of $X$  is masked by generic pre-asymptotic  $A(t)$ such that
$$
 \sigma_X^2(t) = \langle x \rangle (1-\langle x \rangle)\big[(1-g)A(t)+ g t^{\gamma_{X}}\big]
$$ 
 with    $0<g\le1$ and $A(t)$ increasing and such that  $A(t)/t^{\gamma_{X}}\to 0$  for  $t \rightarrow \infty$ (to guarantee that the
 asymptotic behavior is dominated by $t^{\gamma_{X}}$)  and $A(1)=1$ (as $\sigma_X^2(1)=\langle x \rangle (1-\langle x \rangle)$).
 The asymptotic behavior $\sigma^2_Z(t)\sim t^{\gamma_{X}}$ in Eq.(\ref{eq.sumsimple}) dominates only after a time $t_T$ such that: 
\begin{equation}
\frac{\langle\xi\rangle  t_T \,+\,  (1-g)\langle x \rangle (1-\langle \xi\rangle)A(t_T)}{ g(1-\langle\xi\rangle) 
    \langle x \rangle  } =  t_T^{\gamma_{X}}
\end{equation}
Using the fact that  the term $(1-g)\langle x \rangle (1- \langle \xi \rangle) A(t)$ is positive and that $t^{\gamma}$ is monotonically
increasing we finally have
\begin{equation}\label{eq.tt}
t_T\ge t_T^*= \left(\frac{\langle \xi \rangle}{1-\langle \xi \rangle} \frac{1}{g \langle x \rangle}\right)^{1/(\gamma_X-1)},
\end{equation}
which corresponds to Eq.~(7) of the main text. In practice, any finite-time estimate $\hat{\gamma}_X$ is close to the asymptotic $\gamma_X$ only if the estimate  is performed for
  $t\gg t_T$, otherwise $\hat{\gamma}_X<\gamma_X$ ($\hat{\gamma}_X=1$ if $t \ll t_T$). 
\newline

As noted in the main text, if {\bf  z} = {\bf x}+{\bf y}  then  $\bar{\bf x}$ = $\bar{\bf z}$ +{\bf y}. Applying to this relation the same arguments above, similar  pre-asymptotic normal diffusion and  transition time appear in the case of random subtraction, moving up in the hierarchy. More specifically, starting from a sequence {\bf z} such that asymptotically
$\sigma^2_Z (t) \simeq g\langle z \rangle (1- \langle z \rangle) t^{\gamma_Z}$ and constructing ~${\bf x}={\bf \zeta z}$, with {\bf $\zeta$} independent of {\bf z}, we obtain a transition time $t_T$ for ${\bf x}$ given by: 
\begin{equation}\label{eq.ttsub}
t_T\ge t_T^*= \left(\frac{1-\langle \zeta \rangle}{\langle \zeta \rangle} \frac{1}{g (1-\langle z \rangle)}\right)^{1/(\gamma_Z-1)},
\end{equation}
which corresponds to Eq.~(\ref{eq.tt}) above after properly replacing
$\langle x \rangle \to  (1- \langle z\rangle)$, ${\langle \xi \rangle } \to (1-\langle \zeta \rangle)$ and  $\gamma_X \to \gamma_Z$.

\subsection{Random subtraction preserves burstiness}\label{tail-time}

We consider the case of sequences as in Eq. (8) of the main text: $\textbf{z}$ is a sequence emerging from a renewal process with algebraically decaying inter-event times, i.e. $p(\tau )=\tau^{-\mu}$ and $C_\tau(k)=\delta(k)$. Given now a fixed $0\leq \langle \xi \rangle \leq 1$,  
we consider the random subtraction ${\bf x}={\bf \xi} {\bf z}$  where each $z_j=1$ is eventually set to $z_j=0$ with probability $\langle \xi
\rangle$. It is easy to see that the inter-event times of the
new process will be distributed as: 
$$
{\tilde p}(\tau)= (1-\langle \xi \rangle)p(\tau) + \sum_{k\geq 1}^{\infty} (\langle \xi \rangle)^k\sum_{t_1+t_2+\cdots+t_k=\tau} \prod_{j=1}^{k} p(t_j).
$$
Asymptotically ${\tilde p}(\tau)$  is dominated by the long tails of $(1-\langle \xi \rangle)p(\tau)$: given a large $\tau$, fix $\bar{k}>0$ eventually diverging with $\tau\to\infty$ and split accordingly the sum over $k$ in the second term of the right hand side. The term corresponding to the sum $k>\bar{k}$ is exponentially dominated by $\xi^{\bar{k}}$ and arbitrary small, while the remaining  finite sum over $k\leq \bar{k}$ is controlled again by the tail of $p(\tau)$.

\section{Data}\label{sec.data}

In our investigations we considered the English version of the $10$ popular novels listed in SI-Tab. Books. The texts were obtained
through the Gutenberg project (\url{http://www.gutenberg.org}).
We implement a very mild pre-processing of the text that reduces the number of different symbols and simplifies our analysis: we consider as valid symbols the letters
``a-z'', numbers ``0-9'', the apostrophe~ `` ' '' and the blank space ``~''.  Capitalization, punctuations and other markers were
removed. A string of symbols between two consecutive blank spaces is considered to be a word.  No lemmatization was applied to them so
that plurals and singular forms are considered to be different words. 

\section{Confidence interval for determining long-range correlation}\label{ssec.confidence}

As described in the main text, the distinction between long-range
and short-range correlation requires a finite-time estimate~$\hat{\gamma}$ of the asymptotic diffusion exponent~$\gamma$ of the
random-walkers associated to a binary sequence. In practice, this corresponds to 
estimate the tails of the $\sigma^2 \simeq t^\gamma$ relation and it is therefore
essential to estimate the upper limit in~$t$, denoted as~$t_s$, for which we have enough
accuracy to provide a reasonable estimate~$\hat{\gamma}$.   
We adopt the following procedure to estimate~$t_s$. We consider a surrogate binary
sequence with the same length~$N$ and fraction of symbols ($1$'s), but with the
symbols randomly placed in the sequence. For this sequence we know that~$\gamma=1$. We
then consider instants of time~$t_i$ equally spaced in a logarithmic scale of~$t$ (in
practice we consider $t_{i+1}/t_i=1.2$, with $i$ integer and $t_0=1$). We then estimate the local
exponent as
$\hat{\gamma}_{\text{local}}(t_i)=[\log_{10}\Delta\sigma^2(t_{i+1})-\log_{10}\Delta\sigma^2(t_{i})]/\log_{10}(1.2)$. For
small~$t$, $\hat{\gamma}_{\text{local}}=1$ but for larger~$t$ statistical fluctuations arise due
to the finiteness of~$N$, as illustrated in Fig.~\ref{fig.A1}(a). We choose~$t_s$ as the
smallest~$t_i$ for which 
$\{\hat{\gamma}_{\text{local}}(t_{i+1}),\hat{\gamma}_{\text{local}}(t_{i+2}),\hat{\gamma}_{\text{local}}(t_{i+3})\}$
are all outside $[0.9,1.1]$ (see Fig.~\ref{fig.A1}a).  
We recall that our primary interest in the distinction between $\gamma=1$ and~$\gamma\neq1$.   
The procedure described above is particularly suited for this distinction and an exponent~$\hat{\gamma}>1.1$ obtained for 
 large~$t \lessapprox t_s$ can be confidently regarded as a signature of super diffusion
(long-range correlation).   
In Fig.~\ref{fig.A1} we verify that~$t_s$ show no strong dependence on the fraction of
$1$'s in the binary sequence (inset) and that it scales linearly with~$N$. Based on
these results, a good estimate of~$t_s$ is~$t_s=N/100$, i.e. the safe interval for
determining long-range correlation ends two decades before the size of the text. This
phenomenological rule was adopted in the estimate of~$\hat{\gamma}$ for all cases. The~$t_s$ is
only the upper limit and the estimate~$\hat{\gamma}$ is performed through a least-squared
fit in the time interval $t_{s'} < t < t_s=N/100$, where $t_{s'} \approx t_s/100$. In
practice, we select $10$ different values of $t_i$ around $t_s/100$ and report the mean and variance over the different fittings
as~$\hat{\gamma}$ and its uncertainty, respectively.

\section{Lower bound for~$\hat{\gamma}$ due to burstiness}

We start clarifying the validity of the inequality
\begin{equation}\label{eq.inequality}
\hat{\gamma} \ge \hat{\gamma}_{A2},
\end{equation}
where $\hat{\gamma}$ is the finite-time estimate of the total long-range correlation~$\gamma$ of a binary sequence {\bf x} and
$\gamma_{A2}$ is the estimate for the correlation due to the burstiness (which can be quantified by shuffling {\bf x} using the procedure~$A2$
of the main text). 
Equation (4) of the main text shows that both burstiness $\sm \rightarrow \infty$ and long-range correlations in the sequence of $\tau_i$'s
contribute to the long-range correlations of a binary sequence ${\bf x}$.  While the $\sm$ contribution is always positive, the contribution
from the correlation in $\tau_i$'s can be positive or negative. In principle, a negative contribution could precisely cancel the contribution of
$\sm$ and violates the inequality~(\ref{eq.inequality}). 
Conversely, this inequality is guaranteed to hold if the asymptotic contribution of the correlation in $\tau_i$'s of ${\bf x}$ to
$\sigma^2_X$ is positive. 
We now show that this is the case for the sequences we have argued to provide a good account of our observations. 
Consider high in the hierarchy a renewal sequence {\bf x} with a given $\gamma >1$ and broad tail in $p(\tau)$ (diverging $\sm$). Adding
many independent non-overlapping sequences,  we construct a  lower level sequence that  still has long 
range correlation, with the same exponent $\gamma$ (see Sec.~\ref{sec.operations} above). For this sequence we know that  the broad tail in
$p(\tau)$ has a cutoff $\tau_c$
and thus burstiness gives no contribution to $\gamma$. Instead, $\gamma>1$ results solely from the correlations in the {$\tau$}'s, which are
therefore necessarily positive. It is natural to expect that this positiveness of the asymptotic correlation extends to finite times, in
which case the (finite time) inequality~(\ref{eq.inequality}) holds. Indeed, for small $\tau<\tau_c$, the distribution $p(\tau)$ is not
strongly affected by the independent additions and thus for $t<\tau_c$ a finite time estimate $\hat{\gamma}$  will receive contributions
from both burstiness and $\tau's$ correlations.
Finally, we have directly tested the validity of Eq.~(\ref{eq.inequality}) by comparing~$\hat{\gamma}$ of different sequences {\bf x} to the
$\hat{\gamma}_{A2}$ obtained from the corresponding ${\bf x}_{A2}$ (A2-shuffled sequences of {\bf x}, see main text).
The inequality~(\ref{eq.inequality}) was confirmed for every single sequence we have analyzed, as shown by the fact that $\hat{\gamma}_{A2}$
(red symbols) in Fig.~\ref{fig.3si} are systematically below their corresponding $\hat{\gamma}_X$ (black circles). 

We now obtain a quantitative lower bound for~$\hat{\gamma}$ using Eq.~(\ref{eq.inequality}). We consider a renewal sequence (in which case $\hat{\gamma}=\hat{\gamma}_{A2}$)
with an inter-event time distribution given by
\begin{equation}\label{cutoffPT}
p(\tau)=C\tau^{-(4-\gamma_{A2})}e^{-{\frac{\tau}{\tau_c}}}, \qquad \tau>\tau_{min},
\end{equation}
where  $\tau_c$ is the cut-off time, $\gamma_{A2}$ is the anomalous diffusion exponent for a renewal sequence with no cutoff $\tau_c
\rightarrow \infty$, $\tau_{min}$ is a lower cut-off (we fixed it at $\tau_{min}=10$), and $C$ is a normalization constant.
We obtain the lower bound for~$\hat{\gamma}$ as a function of $\sm$ by considering how $\hat{\gamma}_{A2}$ and $\sm$ change with
  $\tau_c$ in the model above.
For short times ($t<<\tau_c$) the corresponding walkers have not seen the cutoff and their diffusion will be anomalous with
exponent $\hat{\gamma}_{A2}=\gamma_{A2}$. At longer time ($t>>\tau_c$~\cite{Mantegna,DelC,Shlesinger}) the diffusion becomes normal
$\hat{\gamma}_{A2}=1$.
Correspondingly, if the fitting interval $t\in [t_s',t_s]$ used to compute the finite time $\hat{\gamma}_{A2}$ (see Sec.~\ref{ssec.confidence}) is all below
$\tau_c$ (i.e. $t_s<\tau_c$)  we have $\hat{\gamma}_{A2}=\gamma_{A2}$ while if the fitting interval is all beyond the cutoff
(i.e. $\tau_c<t_{s'}$) we have $\hat{\gamma}_{A2}=1$. When $\tau_c$ is inside the fitting interval we approximate  $\hat{\gamma}_{A2}$ by 
linearly interpolating between $\gamma_{A2}$ and $1$. Finally, we can compute  $\sm$ by directly calculating the first and second moments of
the distribution~(\ref{cutoffPT}). Particularly important are the values $s_1$ and $s_2$ obtained evaluating $\sm$  at
the critical values of the cutoff $\tau_c=t_{s'}$ and $\tau_c=t_s$, respectively. 
Using the fact that $\sm$ is a monotonic increasing function of $\tau_c$ we can obtain explicitely the $\hat{\gamma}$ dependency on $\sm$. 
The $\hat{\gamma}_{A2}$ for the case of a binary sequence
with distribution~(\ref{cutoffPT}) is given by
\begin{eqnarray*}\label{lowerbound}
\hat{\gamma}_{A2}&=1 \qquad \, \qquad  \qquad   \qquad \qquad & \textrm{if} \qquad \sm < s_1,\\
\hat{\gamma}_{A2}&= (\sm-s_1)\frac{(\gamma_{A2}-1)}{(s_2-s_1)}+1 \qquad & \textrm{if} \qquad \sm \in[ s_1, s_2],\\
\hat{\gamma}_{A2}&=\gamma_{A2} \qquad \qquad \qquad   \qquad  \qquad & \textrm{if} \qquad \sm > s_2.
\end{eqnarray*}
The red dashed line in Fig.~\ref{fig.3si} (Fig. 3 of the main text) was computed using the fitting range corresponding to the book
wrnpc~$t_{s'}=3 \; 10^2, t_{s}= 3 \; 10^4$ (see Sec.~\ref{ssec.confidence}), and $\gamma_{A2}=1.6$ (compatible with $\hat{\gamma}$ observed
for words with large $\sm$).

\section{Additional shuffling methods}

In addition to the shuffling methods presented in the main text, we discuss here briefly two cases: 
\begin{itemize}
\item  {\bf Shuffle words}\\
Mixing words order kills correlations for scales larger than the maximum word length~\cite{ebeling3,monte}.
Even the blank space sequence~${\bf B}$ becomes uncorrelated 
because its original correlations originate (as in the case of
all letters) from the correlation in~$\tau_i$ and not from tails in~$p(\tau)$. 
\item  {\bf Keep all blank spaces} in their original positions and fill the empty space between them with:
\begin{itemize}
\item [1-] two letters $a,b$, placed randomly with probabilities $p_a=p$ and $p_b=1-p$.
\item [2-] the same letters of the book, placed in random positions.
\end{itemize}
By construction, 
correlation for blank space is trivially preserved. What do we expect for the other letters? The following simple reasoning indicates that long-range
correlation should be expected asymptotically in both cases: any letter sequence ${\bf x}$ is on top of the reverted blank space sequence ${\bf \bar{B}}$; 
the results in Sec.~\ref{sec.operations} above show that either the selected
sequence ${\bf x}$ or its complement~${\bf y}$ (such that ${\bf x}+{\bf y}=\bar{{\bf B}}$) has~$\gamma=\gamma_B$; and Eq.~(\ref{eq.sumsimple})
above shows that any randomly chosen ${\bf x}$ on top of ${\bf \bar{B}}$ has~$\gamma=\gamma_B$. In practice these exponents are relevant only if the 
subsequence is dense enough in order for $t_T$ in Eq.~(\ref{eq.ttsub}) above to be inside the observation range. For the  first shuffling method and  for 
our longest book (wrnpc), we obtain that only if $p>95.8\%$ one finds $t_T<t_s=1\%$ book size. Since the most
frequent letter in a book has much smaller frequency (around $10\%$), we conclude that in practice all sequences obtained using the second
shuffling mehthod have $\hat{\gamma}=1$ for all books of size smaller than~$100 \times t_T\approx10^{11}$ symbols ($\approx 10^7$ pages). 

These simple calculations show that $\gamma_B>1$ does not explain the correlations
observed in the letters of the original text, as has been speculated in Ref.~\cite{ebeling1}. Their origin are the long-range correlations
on higher levels.  
\end{itemize}


\setcounter{figure}{0}
\renewcommand{\thefigure}{S\arabic{figure}}
\renewcommand\figurename{{\bf Fig.}}

\begin{figure*}[!ht]
\includegraphics[width=1.5\columnwidth]{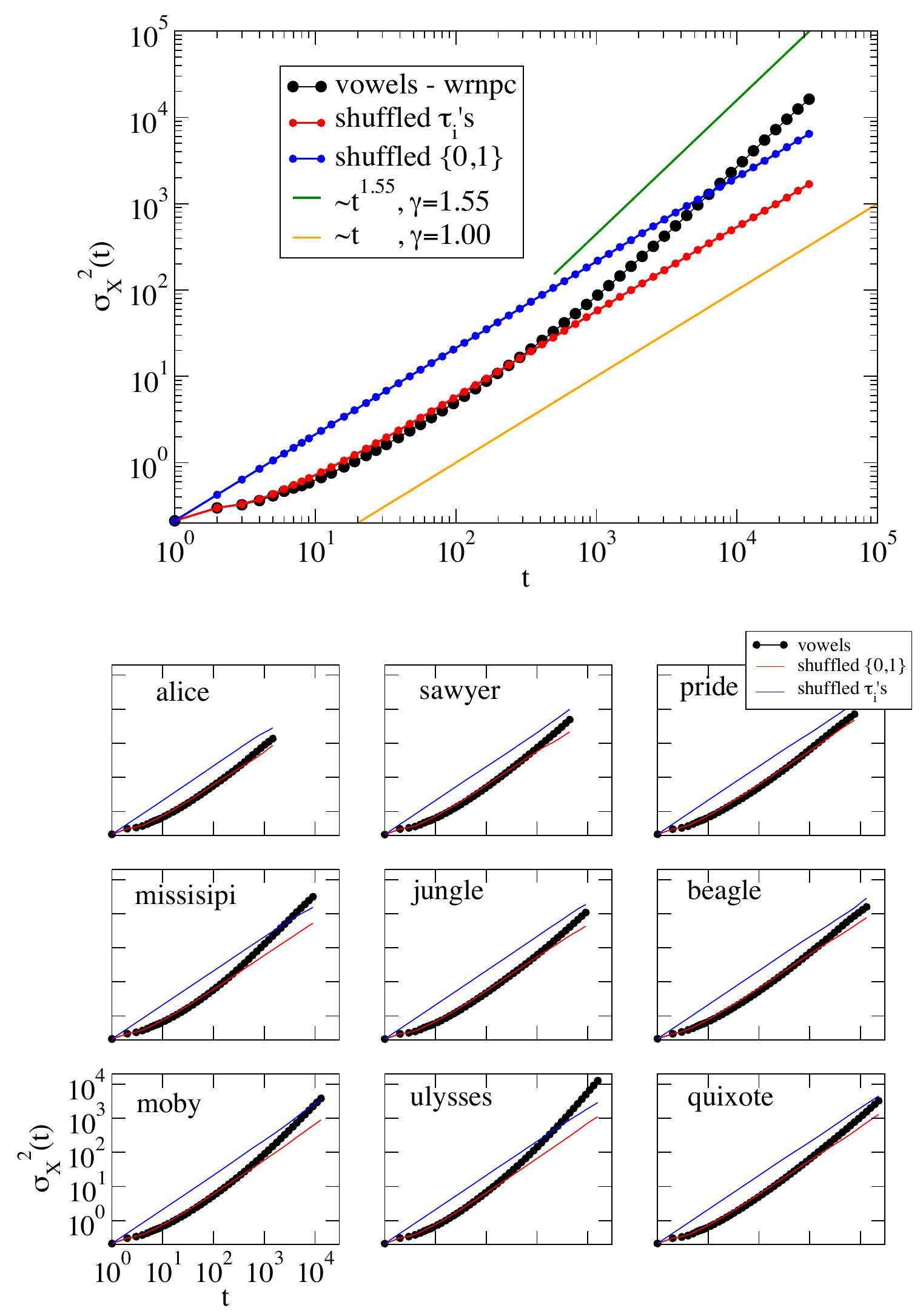}
\caption{Long-range correlation in texts encoded as vowels.  Upper plot: detailed analysis in the book wrnpc with
  exponent~$\hat{\gamma} =1.55 \pm 0.05$ (wrnpc). Lower plots: analysis of the remaining $9$ books with the following exponents~$\hat{\gamma}$: $1.55
  \pm 0.05$ (wrnpc), $1.18 \pm 0.05$ (alice), $1.23 \pm 0.04$ (sawyer), $1.20 \pm 0.03$ (pride)
$1.48 \pm 0.05 $ (missisipi), $1.26\pm0.05$ (jungle), $1.25 \pm 0.04$ (beagle), $1.45 \pm 0.05$ (moby),
$1.61 \pm 0.06$ (ulysses), $1.26 \pm 0.04$ (quixote)}\label{fig.vowels9} 
\end{figure*}

\begin{figure*}[!ht]
\includegraphics[width=1.8\columnwidth]{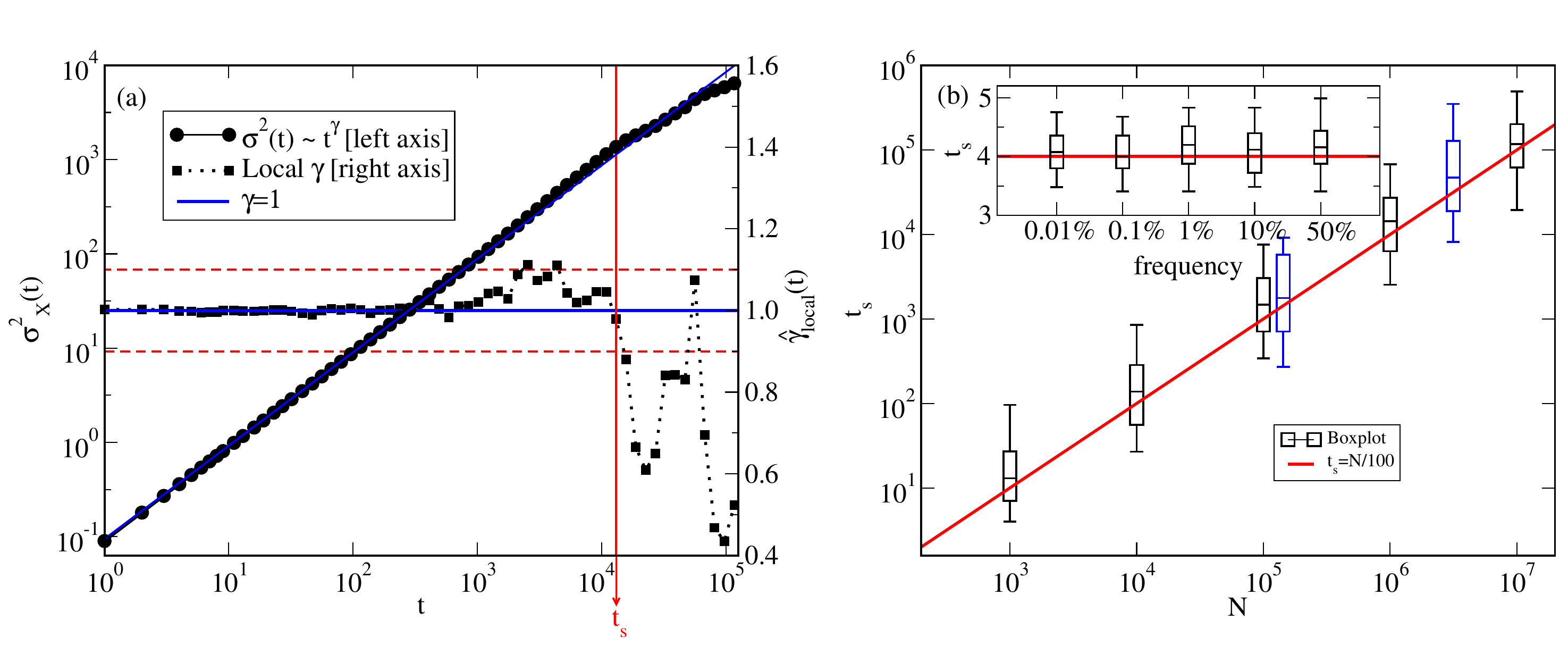}
\caption{ (Color online) Determination of the time interval for the estimate of the long-range correlation exponent~$\hat{\gamma}$. (a)
  The dispersion~$\Delta \sigma^2$ as a function of time~$t$ is shown as $\newmoon$ for a random binary sequence of size $N=10^6$ and $10\%$ of $1$'s. The local derivative is shown as $\blacksquare$ and agrees with the theoretical exponent~$\gamma=1$ until fluctuations start for long~$t$ (axis on the right). The time~$t_s$ denotes the end of the interval of safe determination of~$\gamma$, as explained in the text. (b) Dependence of~$t_s$ on the size of the binary sequence~$N$. The boxplots show the $5\%,25\%,50\%$ (median),$75
\%,$ and~$95\%$ quantiles over $M$ different realizations of a random binary sequence. Black boxplots:~$M=300$ ($M=44$ for $N=10^7$)
realizations equally divided between frequency$=1\%,10\%,50\%$. Blue boxplots: $M=35$ realizations equally divided between the frequencies
of the three most frequent letters  (``$\_$'',``e'', ``t'') and two most frequent words (``$\_$the$\_$",``$\_$and$\_$") of the shortest (Alice, $N=143,488$) and longest (War and Peace, $N=3,147,284$) books
Inset: boxplots of $t_s$ for different frequencies and fixed sequence length~$N=10^6$ and $M=100$, showing no strong dependence on frequency.}\label{fig.A1}
\end{figure*}

\begin{figure*}[!ht]
\includegraphics[width=1.3\columnwidth]{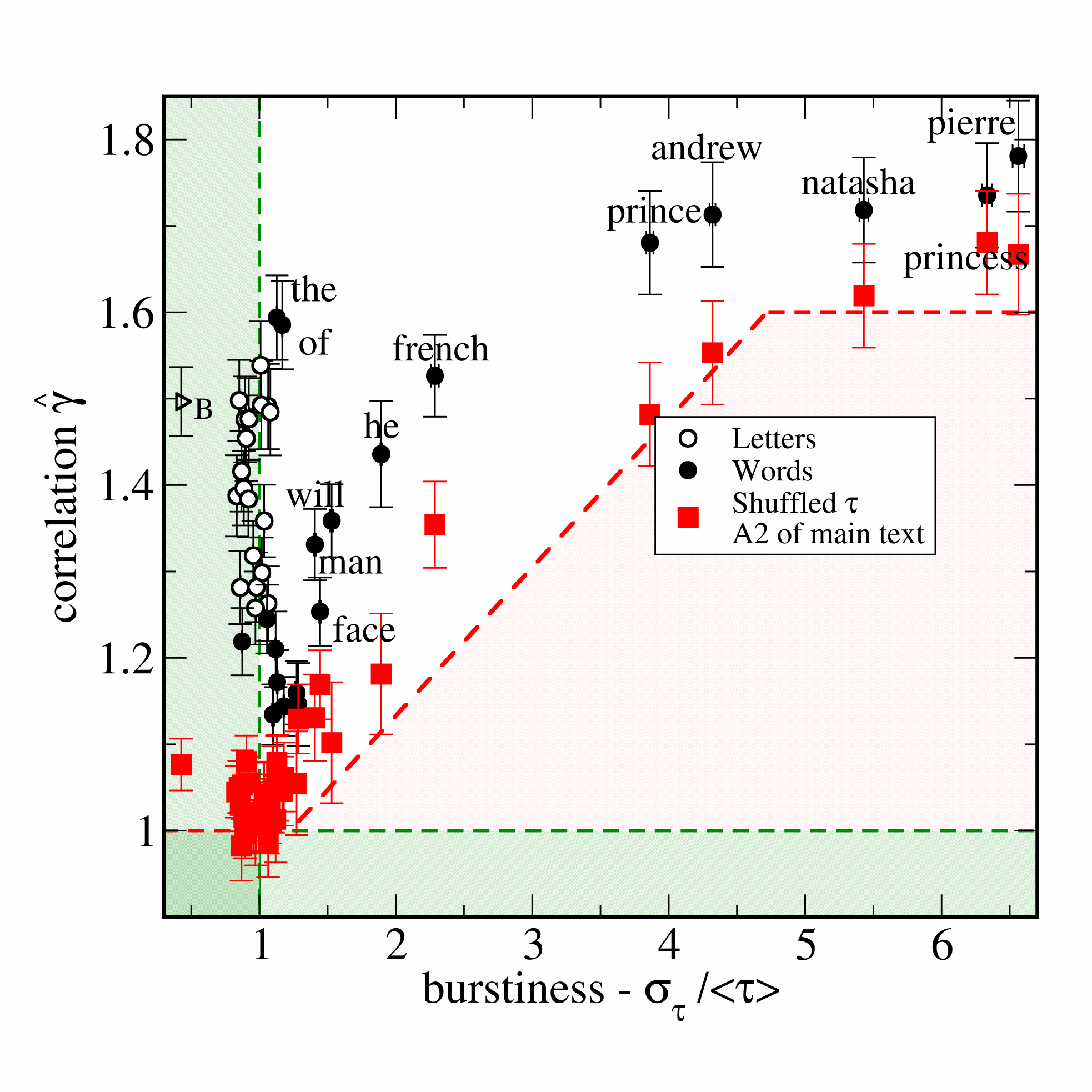}
\caption{This figure corresponds to Fig.~3 of the main paper with the addition (red squares) of the estimated $\hat{\gamma}$ for sequences ${\bf x}_{A2}$
  obtained shuffling each one of the original sequences. The shuffling does not change the $\sm$ and therefore the original and shuffled
  sequences appear always on the same vertical line. The fact that the results for ${\bf x}_{A2}$ are systematically {\it below}  their
  corresponding ${\bf x}$ is a strong evidence of the validity of the inequality~(\ref{eq.inequality}).}\label{fig.3si}
\end{figure*}

\newpage

\setcounter{table}{0}
\renewcommand{\thetable}{S\arabic{table}}
\renewcommand\tablename{{\bf Table}}

\begin{sidewaystable}
 \begin{tabular}{|c|c|c|c|c|c|}
\hline
Short name  & $N$ - Number of Symbols & Title & Author  & Translator and Information from Project Gutenberg\\
\hline
\hline
alice	&	134,847	&	Alice's Adventures in Wonderland	& Lewis Carroll & Released: 2009-05-19 \\
\hline
sawyer	&	369,222		&The Adventures of Tom Sawyer	& Mark Twain &  Released: 2006-08-20\\
\hline
pride	&	659,408		&Pride and Prejudice	& Jane Austen &  Updated: 2010-09-05; Released: June, 1998 \\
\hline
missisipi&	772,391		&Life On The Mississippi  & Mark Twain &  Released: 2004-08-20\\
\hline
jungle	&	783,014		&The Jungle  		& Upton Sinclair &  Release Date: 2006-03-11 \\
\hline
beagle	&	1,153,638		&The Voyage Of The Beagle & Charles Darwin &  Released: February 2003; Reprint from: June 1913 \\
\hline
moby	&	1,169,850		&Moby Dick; or The Whale  &Herman Melville & Updated: 2009-01-03; Released: June 2003 \\
\hline
ulysses	&	1,453,586		&Ulysses	      	     	&James Joyce &  Released: July, 2003 \\
\hline
quixote	&	2,080,431		&Don Quixote		&Miguel de Cervantes Saavedra & Translator: John Ormsby \\
&	&  & &  Released: 2004-07-27 \\
\hline
wrnpc	&	3,082,079		&War and Peace			&Leo Tolstoy & Translator: Louise and Aylmer Maude \\ 
&	&  & & Updated: 2007-05-07; Released: April 2001 \\
\hline
\end{tabular}
\caption{List of books considered in our investigations. The texts were retrieved from the Project Gutenberg {\it www.gutenberg.org} on 21-09-2010} 
\end{sidewaystable}

\begin{table}[ht]
\caption{Correlation~$\hat{\gamma}$ and burstiness $\sigma_\tau/\langle \tau \rangle$ obtained for the diferrent binary sequences in the indicated book. }
 \begin{tabular}{|c||c|c|c|c|c||c|c|c|c|}
\hline
 \multicolumn{10}{|c|}{Book: alice; N=134,847} \\
\hline
\hline
\multicolumn{2}{|c|}{} & \multicolumn{4}{|c|}{Original data  }   & \multicolumn{2}{|c|}{Shuffling  M1} & \multicolumn{2}{|c|}{Shuffling  M2} \\
\hline
sequence & $N_i$ & $\sigma_\tau/\langle \tau \rangle$ & error & $\hat{\gamma}$ & error & $\hat{\gamma}$ & error & $\hat{\gamma}$ & error\\ 
\hline
vowels & 41414 & 0.440 & 0.020 & 1.18 & 0.05 &  &  &  & \\ 
\_ & 26666 & 0.379 & 0.011 & 1.13 & 0.06 & 1.13 & 0.06 & 1.13 & 0.06\\ 
e & 13545 & 0.812 & 0.003 & 1.20 & 0.04 & 1.11 & 0.04 & 1.01 & 0.04\\ 
t & 10667 & 0.858 & 0.003 & 1.17 & 0.05 & 1.05 & 0.03 & 1.05 & 0.03\\ 
a & 8772 & 0.838 & 0.003 & 1.14 & 0.05 & 1.07 & 0.03 & 0.98 & 0.04\\ 
o & 8128 & 0.920 & 0.002 & 1.25 & 0.05 & 1.13 & 0.04 & 0.99 & 0.04\\ 
i & 7500 & 0.887 & 0.002 & 1.20 & 0.04 & 1.10 & 0.03 & 1.03 & 0.03\\ 
h & 7379 & 0.848 & 0.003 & 1.15 & 0.04 & 1.11 & 0.04 & 1.04 & 0.03\\ 
n & 7001 & 0.895 & 0.002 & 1.09 & 0.03 & 1.13 & 0.04 & 1.02 & 0.03\\ 
s & 6497 & 0.925 & 0.002 & 1.11 & 0.04 & 1.09 & 0.03 & 1.07 & 0.03\\ 
r & 5418 & 0.905 & 0.002 & 1.15 & 0.04 & 1.15 & 0.04 & 1.04 & 0.03\\ 
d & 4928 & 0.878 & 0.003 & 1.06 & 0.03 & 1.10 & 0.04 & 0.97 & 0.04\\ 
l & 4704 & 1.081 & 0.003 & 1.20 & 0.06 & 1.12 & 0.04 & 1.00 & 0.03\\ 
u & 3469 & 0.901 & 0.004 & 1.15 & 0.04 & 1.15 & 0.04 & 1.07 & 0.03\\ 
w & 2681 & 0.966 & 0.003 & 1.11 & 0.04 & 1.23 & 0.05 & 0.99 & 0.04\\ 
g & 2529 & 0.986 & 0.003 & 1.13 & 0.04 & 1.16 & 0.05 & 0.97 & 0.05\\ 
c & 2397 & 0.980 & 0.005 & 1.15 & 0.05 & 1.11 & 0.04 & 1.00 & 0.03\\ 
y & 2259 & 1.070 & 0.004 & 1.23 & 0.04 & 1.05 & 0.03 & 1.00 & 0.04\\ 
m & 2103 & 1.030 & 0.005 & 1.16 & 0.04 & 1.24 & 0.05 & 0.98 & 0.03\\ 
f & 1988 & 1.089 & 0.006 & 1.17 & 0.05 & 1.14 & 0.04 & 1.02 & 0.04\\ 
p & 1514 & 1.143 & 0.006 & 1.17 & 0.04 & 1.13 & 0.04 & 1.05 & 0.03\\ 
\_the\_ & 1635 & 0.971 & 0.005 & 1.29 & 0.07 &  &  &  & \\ 
\_and\_ & 868 & 0.973 & 0.008 & 1.08 & 0.04 &  &  &  & \\ 
\_to\_ & 734 & 1.013 & 0.006 & 1.08 & 0.04 &  &  &  & \\ 
\_a\_ & 624 & 1.057 & 0.007 & 1.08 & 0.03 &  &  &  & \\ 
\_she\_ & 542 & 1.548 & 0.012 & 1.34 & 0.06 &  &  &  & \\ 
\_it\_ & 530 & 1.172 & 0.008 & 1.22 & 0.04 &  &  &  & \\ 
\_alice\_ & 386 & 0.885 & 0.008 & 1.01 & 0.05 &  &  &  & \\ 
\_in\_ & 367 & 0.959 & 0.008 & 1.04 & 0.03 &  &  &  & \\ 
\_way\_ & 57 & 1.098 & 0.021 & 1.21 & 0.04 &  &  &  & \\ 
\_turtle\_ & 57 & 4.066 & 0.039 & 1.47 & 0.12 &  &  &  & \\ 
\_hatter\_ & 55 & 4.978 & 0.050 & 1.46 & 0.12 &  &  &  & \\ 
\_gryphon\_ & 55 & 3.541 & 0.038 & 1.43 & 0.10 &  &  &  & \\ 
\_quite\_ & 55 & 1.290 & 0.025 & 1.16 & 0.04 &  &  &  & \\ 
\_mock\_ & 55 & 3.919 & 0.045 & 1.45 & 0.12 &  &  &  & \\ 
\_are\_ & 54 & 1.250 & 0.024 & 1.10 & 0.03 &  &  &  & \\ 
\_think\_ & 52 & 1.268 & 0.039 & 1.09 & 0.04 &  &  &  & \\ 
\_more\_ & 49 & 1.040 & 0.023 & 1.11 & 0.04 &  &  &  & \\ 
\_head\_ & 49 & 1.230 & 0.024 & 1.07 & 0.03 &  &  &  & \\ 
\_never\_ & 48 & 1.083 & 0.049 & 1.09 & 0.05 &  &  &  & \\ 
\_voice\_ & 47 & 1.359 & 0.054 & 1.07 & 0.03 &  &  &  & \\ 
\hline
\end{tabular}
\end{table}

\begin{table}[ht]
\caption{Correlation~$\hat{\gamma}$ and burstiness $\sigma_\tau/\langle \tau \rangle$ obtained for the diferrent binary sequences in the indicated book. }
 \begin{tabular}{|c||c|c|c|c|c||c|c|c|c|}
\hline
 \multicolumn{10}{|c|}{Book: beagle; N=1,153,638 }\\
\hline
\hline
\multicolumn{2}{|c|}{} & \multicolumn{4}{|c|}{Original data  }   & \multicolumn{2}{|c|}{Shuffling  M1} & \multicolumn{2}{|c|}{Shuffling  M2} \\
\hline
sequence & $N_i$ & $\sigma_\tau/\langle \tau \rangle$ & error & $\hat{\gamma}$ & error & $\hat{\gamma}$ & error & $\hat{\gamma}$ & error\\ 
\hline
vowels & 358397 & 0.454 & 0.020 & 1.25 & 0.04 &  &  &  & \\ 
\_ & 208375 & 0.434 & 0.009 & 1.34 & 0.04 & 1.34 & 0.04 & 1.34 & 0.04\\ 
e & 123056 & 0.817 & 0.002 & 1.18 & 0.04 & 1.22 & 0.05 & 0.96 & 0.05\\ 
t & 86576 & 0.836 & 0.002 & 1.18 & 0.04 & 1.15 & 0.04 & 0.98 & 0.04\\ 
a & 78914 & 0.847 & 0.002 & 1.11 & 0.04 & 1.11 & 0.04 & 1.03 & 0.03\\ 
o & 67896 & 0.885 & 0.001 & 1.20 & 0.04 & 1.24 & 0.04 & 0.96 & 0.05\\ 
n & 64597 & 0.865 & 0.002 & 1.13 & 0.04 & 1.18 & 0.04 & 1.02 & 0.04\\ 
i & 63755 & 0.889 & 0.001 & 1.22 & 0.04 & 1.15 & 0.04 & 1.07 & 0.03\\ 
s & 62383 & 0.909 & 0.001 & 1.21 & 0.04 & 1.20 & 0.04 & 1.05 & 0.03\\ 
r & 59027 & 0.870 & 0.001 & 1.17 & 0.04 & 1.09 & 0.04 & 1.05 & 0.03\\ 
h & 54880 & 0.861 & 0.002 & 1.23 & 0.04 & 1.07 & 0.05 & 1.09 & 0.03\\ 
l & 38467 & 1.033 & 0.001 & 1.28 & 0.05 & 1.10 & 0.04 & 1.03 & 0.03\\ 
d & 37051 & 0.923 & 0.001 & 1.24 & 0.04 & 1.19 & 0.04 & 1.01 & 0.03\\ 
c & 27687 & 0.978 & 0.001 & 1.28 & 0.04 & 1.17 & 0.04 & 1.11 & 0.03\\ 
u & 24776 & 0.957 & 0.001 & 1.15 & 0.04 & 1.12 & 0.04 & 1.04 & 0.03\\ 
f & 24052 & 0.955 & 0.001 & 1.18 & 0.04 & 1.12 & 0.04 & 1.08 & 0.03\\ 
m & 21509 & 0.987 & 0.001 & 1.20 & 0.04 & 1.20 & 0.04 & 1.03 & 0.04\\ 
w & 19172 & 1.010 & 0.001 & 1.36 & 0.05 & 1.21 & 0.04 & 1.07 & 0.03\\ 
g & 18284 & 0.962 & 0.001 & 1.20 & 0.04 & 1.09 & 0.05 & 0.99 & 0.03\\ 
p & 16742 & 1.040 & 0.001 & 1.24 & 0.04 & 1.10 & 0.03 & 1.04 & 0.03\\ 
y & 15700 & 0.993 & 0.002 & 1.26 & 0.04 & 1.15 & 0.04 & 1.03 & 0.04\\ 
\_the\_ & 16882 & 0.924 & 0.002 & 1.21 & 0.04 &  &  &  & \\ 
\_of\_ & 9414 & 0.970 & 0.002 & 1.25 & 0.04 &  &  &  & \\ 
\_and\_ & 5765 & 0.897 & 0.003 & 1.10 & 0.04 &  &  &  & \\ 
\_a\_ & 5326 & 1.097 & 0.003 & 1.19 & 0.04 &  &  &  & \\ 
\_in\_ & 4287 & 1.022 & 0.003 & 1.14 & 0.04 &  &  &  & \\ 
\_to\_ & 4080 & 1.051 & 0.003 & 1.20 & 0.04 &  &  &  & \\ 
\_water\_ & 417 & 1.509 & 0.011 & 1.26 & 0.04 &  &  &  & \\ 
\_little\_ & 412 & 1.117 & 0.011 & 1.08 & 0.03 &  &  &  & \\ 
\_where\_ & 349 & 1.086 & 0.011 & 1.06 & 0.03 &  &  &  & \\ 
\_sea\_ & 348 & 1.534 & 0.015 & 1.29 & 0.04 &  &  &  & \\ 
\_much\_ & 338 & 1.112 & 0.010 & 1.08 & 0.04 &  &  &  & \\ 
\_country\_ & 337 & 1.519 & 0.011 & 1.28 & 0.05 &  &  &  & \\ 
\_land\_ & 318 & 1.387 & 0.012 & 1.25 & 0.04 &  &  &  & \\ 
\_must\_ & 317 & 1.290 & 0.009 & 1.12 & 0.04 &  &  &  & \\ 
\_feet\_ & 312 & 1.391 & 0.013 & 1.23 & 0.04 &  &  &  & \\ 
\_may\_ & 311 & 1.118 & 0.010 & 1.09 & 0.03 &  &  &  & \\ 
\_species\_ & 303 & 2.459 & 0.022 & 1.55 & 0.05 &  &  &  & \\ 
\_found\_ & 303 & 1.217 & 0.010 & 1.16 & 0.04 &  &  &  & \\ 
\_me\_ & 301 & 1.206 & 0.012 & 1.07 & 0.04 &  &  &  & \\ 
\_day\_ & 301 & 1.375 & 0.007 & 1.12 & 0.03 &  &  &  & \\ 
\hline
\end{tabular}
\end{table}

\begin{table}[ht]
\caption{Correlation~$\hat{\gamma}$ and burstiness $\sigma_\tau/\langle \tau \rangle$ obtained for the diferrent binary sequences in the indicated book. }
 \begin{tabular}{|c||c|c|c|c|c||c|c|c|c|}
\hline
 \multicolumn{10}{|c|}{Book: jungle; N=783,014 }\\
\hline
\hline
\multicolumn{2}{|c|}{} & \multicolumn{4}{|c|}{Original data}   & \multicolumn{2}{|c|}{Shuffling  M1} & \multicolumn{2}{|c|}{Shuffling  M2} \\
\hline
sequence & $N_i$ & $\sigma_\tau/\langle \tau \rangle$ & error & $\hat{\gamma}$ & error & $\hat{\gamma}$ & error & $\hat{\gamma}$ & error\\ 
\hline
vowels & 237050 & 0.420 & 0.020 & 1.26 & 0.05 &  &  &  & \\ 
\_ & 151300 & 0.404 & 0.010 & 1.53 & 0.05 & 1.53 & 0.05 & 1.53 & 0.05\\ 
e & 78161 & 0.843 & 0.002 & 1.16 & 0.04 & 1.09 & 0.04 & 1.04 & 0.04\\ 
t & 58475 & 0.873 & 0.002 & 1.32 & 0.05 & 1.19 & 0.04 & 1.04 & 0.03\\ 
a & 53663 & 0.854 & 0.002 & 1.21 & 0.04 & 1.17 & 0.04 & 1.08 & 0.03\\ 
o & 47726 & 0.896 & 0.001 & 1.18 & 0.04 & 1.24 & 0.04 & 0.97 & 0.04\\ 
n & 44497 & 0.874 & 0.002 & 1.14 & 0.04 & 1.18 & 0.04 & 1.12 & 0.03\\ 
h & 44473 & 0.832 & 0.002 & 1.37 & 0.05 & 1.27 & 0.04 & 1.17 & 0.05\\ 
i & 40025 & 0.906 & 0.001 & 1.34 & 0.05 & 1.23 & 0.05 & 1.13 & 0.04\\ 
s & 37500 & 0.941 & 0.001 & 1.32 & 0.05 & 1.37 & 0.04 & 1.07 & 0.03\\ 
r & 34514 & 0.888 & 0.002 & 1.19 & 0.04 & 1.19 & 0.04 & 1.09 & 0.04\\ 
d & 30491 & 0.929 & 0.001 & 1.31 & 0.06 & 1.22 & 0.04 & 1.00 & 0.04\\ 
l & 24876 & 1.059 & 0.001 & 1.20 & 0.04 & 1.10 & 0.04 & 1.07 & 0.03\\ 
u & 17475 & 0.943 & 0.001 & 1.16 & 0.04 & 1.22 & 0.04 & 0.99 & 0.05\\ 
w & 17213 & 0.948 & 0.002 & 1.36 & 0.06 & 1.30 & 0.05 & 1.06 & 0.03\\ 
m & 14754 & 0.977 & 0.001 & 1.18 & 0.04 & 1.24 & 0.04 & 1.04 & 0.03\\ 
c & 14148 & 1.006 & 0.001 & 1.37 & 0.05 & 1.18 & 0.04 & 1.10 & 0.04\\ 
g & 14069 & 0.994 & 0.002 & 1.28 & 0.06 & 1.27 & 0.04 & 1.06 & 0.03\\ 
f & 13862 & 1.016 & 0.002 & 1.25 & 0.04 & 1.21 & 0.04 & 1.09 & 0.04\\ 
y & 10868 & 1.068 & 0.002 & 1.29 & 0.05 & 1.16 & 0.04 & 0.95 & 0.05\\ 
p & 9940 & 1.074 & 0.002 & 1.24 & 0.05 & 1.24 & 0.04 & 1.06 & 0.04\\ 
\_the\_ & 8930 & 1.018 & 0.003 & 1.34 & 0.04 &  &  &  & \\ 
\_and\_ & 7280 & 0.958 & 0.002 & 1.27 & 0.04 &  &  &  & \\ 
\_of\_ & 4365 & 1.113 & 0.003 & 1.42 & 0.07 &  &  &  & \\ 
\_to\_ & 4190 & 1.077 & 0.003 & 1.20 & 0.04 &  &  &  & \\ 
\_a\_ & 4158 & 1.152 & 0.004 & 1.22 & 0.04 &  &  &  & \\ 
\_he\_ & 3311 & 2.158 & 0.011 & 1.60 & 0.05 &  &  &  & \\ 
\_him\_ & 1184 & 2.009 & 0.013 & 1.42 & 0.05 &  &  &  & \\ 
\_jurgis\_ & 1098 & 2.077 & 0.010 & 1.48 & 0.07 &  &  &  & \\ 
\_i\_ & 485 & 6.141 & 0.275 & 1.54 & 0.06 &  &  &  & \\ 
\_man\_ & 463 & 1.301 & 0.013 & 1.27 & 0.04 &  &  &  & \\ 
\_said\_ & 367 & 1.975 & 0.019 & 1.38 & 0.04 &  &  &  & \\ 
\_time\_ & 356 & 1.209 & 0.013 & 1.15 & 0.04 &  &  &  & \\ 
\_men\_ & 329 & 1.768 & 0.011 & 1.33 & 0.05 &  &  &  & \\ 
\_now\_ & 325 & 1.077 & 0.009 & 1.11 & 0.03 &  &  &  & \\ 
\_day\_ & 280 & 1.378 & 0.021 & 1.15 & 0.04 &  &  &  & \\ 
\_other\_ & 279 & 1.244 & 0.014 & 1.16 & 0.04 &  &  &  & \\ 
\_place\_ & 263 & 1.227 & 0.013 & 1.17 & 0.04 &  &  &  & \\ 
\_only\_ & 261 & 1.042 & 0.010 & 1.03 & 0.04 &  &  &  & \\ 
\_before\_ & 235 & 1.117 & 0.010 & 1.09 & 0.03 &  &  &  & \\ 
\_home\_ & 229 & 1.759 & 0.012 & 1.23 & 0.04 &  &  &  & \\ 
\hline
\end{tabular}
\end{table}

\begin{table}[ht]
\caption{Correlation~$\hat{\gamma}$ and burstiness $\sigma_\tau/\langle \tau \rangle$ obtained for the diferrent binary sequences in the indicated book. }
 \begin{tabular}{|c||c|c|c|c|c||c|c|c|c|}
\hline
 \multicolumn{10}{|c|}{Book: missisipi; N=772,391}\\
\hline
\hline
\multicolumn{2}{|c|}{} & \multicolumn{4}{|c|}{Original data  }   & \multicolumn{2}{|c|}{Shuffling  M1} & \multicolumn{2}{|c|}{Shuffling  M2} \\
\hline
sequence & $N_i$ & $\sigma_\tau/\langle \tau \rangle$ & error & $\hat{\gamma}$ & error & $\hat{\gamma}$ & error & $\hat{\gamma}$ & error\\ 
\hline
vowels & 235370 & 0.445 & 0.020 & 1.48 & 0.05 &  &  &  & \\ 
\_ & 146786 & 0.429 & 0.009 & 1.65 & 0.05 & 1.65 & 0.05 & 1.65 & 0.05\\ 
e & 76483 & 0.850 & 0.002 & 1.40 & 0.05 & 1.11 & 0.04 & 1.10 & 0.03\\ 
t & 59660 & 0.858 & 0.002 & 1.24 & 0.04 & 1.18 & 0.04 & 1.02 & 0.04\\ 
a & 51642 & 0.859 & 0.002 & 1.23 & 0.04 & 1.18 & 0.04 & 1.11 & 0.04\\ 
o & 47123 & 0.890 & 0.002 & 1.23 & 0.04 & 1.22 & 0.04 & 1.05 & 0.04\\ 
n & 44064 & 0.869 & 0.002 & 1.24 & 0.04 & 1.24 & 0.04 & 1.08 & 0.03\\ 
i & 42750 & 0.920 & 0.001 & 1.30 & 0.04 & 1.19 & 0.04 & 1.11 & 0.03\\ 
s & 38995 & 0.940 & 0.001 & 1.34 & 0.06 & 1.23 & 0.04 & 1.20 & 0.04\\ 
h & 36904 & 0.859 & 0.002 & 1.40 & 0.05 & 1.13 & 0.04 & 1.20 & 0.04\\ 
r & 35465 & 0.912 & 0.001 & 1.34 & 0.05 & 1.19 & 0.04 & 1.16 & 0.04\\ 
d & 27682 & 0.974 & 0.001 & 1.40 & 0.06 & 1.24 & 0.04 & 1.05 & 0.03\\ 
l & 24910 & 1.055 & 0.001 & 1.20 & 0.04 & 1.12 & 0.04 & 1.06 & 0.03\\ 
u & 17372 & 0.947 & 0.002 & 1.20 & 0.04 & 1.17 & 0.04 & 1.07 & 0.03\\ 
w & 15554 & 0.996 & 0.002 & 1.30 & 0.04 & 1.25 & 0.04 & 1.10 & 0.03\\ 
m & 14940 & 1.006 & 0.002 & 1.29 & 0.04 & 1.27 & 0.04 & 1.09 & 0.04\\ 
c & 14884 & 1.042 & 0.001 & 1.35 & 0.05 & 1.21 & 0.04 & 1.21 & 0.06\\ 
f & 14234 & 1.006 & 0.001 & 1.24 & 0.05 & 1.14 & 0.04 & 1.04 & 0.03\\ 
g & 12890 & 1.044 & 0.001 & 1.26 & 0.04 & 1.17 & 0.04 & 1.09 & 0.03\\ 
y & 11994 & 1.022 & 0.002 & 1.34 & 0.04 & 1.14 & 0.04 & 1.02 & 0.03\\ 
p & 11087 & 1.093 & 0.002 & 1.30 & 0.05 & 1.17 & 0.04 & 1.16 & 0.05\\ 
\_the\_ & 9091 & 1.043 & 0.003 & 1.38 & 0.04 &  &  &  & \\ 
\_and\_ & 5898 & 0.995 & 0.003 & 1.34 & 0.05 &  &  &  & \\ 
\_of\_ & 4380 & 1.033 & 0.003 & 1.32 & 0.05 &  &  &  & \\ 
\_a\_ & 4057 & 1.098 & 0.003 & 1.22 & 0.04 &  &  &  & \\ 
\_to\_ & 3545 & 1.095 & 0.004 & 1.24 & 0.04 &  &  &  & \\ 
\_in\_ & 2555 & 1.031 & 0.004 & 1.14 & 0.04 &  &  &  & \\ 
\_would\_ & 480 & 1.552 & 0.012 & 1.26 & 0.04 &  &  &  & \\ 
\_river\_ & 478 & 2.176 & 0.014 & 1.43 & 0.06 &  &  &  & \\ 
\_water\_ & 242 & 1.899 & 0.015 & 1.38 & 0.05 &  &  &  & \\ 
\_she\_ & 239 & 2.055 & 0.022 & 1.44 & 0.06 &  &  &  & \\ 
\_boat\_ & 212 & 1.921 & 0.028 & 1.32 & 0.05 &  &  &  & \\ 
\_here\_ & 210 & 1.508 & 0.015 & 1.24 & 0.04 &  &  &  & \\ 
\_night\_ & 177 & 1.609 & 0.012 & 1.30 & 0.05 &  &  &  & \\ 
\_can\_ & 177 & 1.392 & 0.015 & 1.13 & 0.04 &  &  &  & \\ 
\_go\_ & 176 & 1.275 & 0.010 & 1.16 & 0.04 &  &  &  & \\ 
\_head\_ & 175 & 1.612 & 0.017 & 1.41 & 0.06 &  &  &  & \\ 
\_pilot\_ & 172 & 2.652 & 0.047 & 1.40 & 0.05 &  &  &  & \\ 
\_long\_ & 172 & 1.246 & 0.013 & 1.06 & 0.03 &  &  &  & \\ 
\_first\_ & 164 & 1.132 & 0.018 & 1.11 & 0.04 &  &  &  & \\ 
\_miles\_ & 162 & 1.816 & 0.030 & 1.49 & 0.05 &  &  &  & \\ 
\hline
\end{tabular}
\end{table}

\begin{table}[ht]
\caption{Correlation~$\hat{\gamma}$ and burstiness $\sigma_\tau/\langle \tau \rangle$ obtained for the diferrent binary sequences in the indicated book. }
 \begin{tabular}{|c||c|c|c|c|c||c|c|c|c|}
\hline
 \multicolumn{10}{|c|}{Book: moby; N=1,169,850 }\\
\hline
\hline
\multicolumn{2}{|c|}{} & \multicolumn{4}{|c|}{Original data  }   & \multicolumn{2}{|c|}{Shuffling  M1} & \multicolumn{2}{|c|}{Shuffling  M2} \\
\hline
sequence & $N_i$ & $\sigma_\tau/\langle \tau \rangle$ & error & $\hat{\gamma}$ & error & $\hat{\gamma}$ & error & $\hat{\gamma}$ & error\\ 
\hline
vowels & 356037 & 0.441 & 0.020 & 1.45 & 0.05 &  &  &  & \\ 
\_ & 215939 & 0.424 & 0.009 & 1.54 & 0.05 & 1.54 & 0.05 & 1.54 & 0.05\\ 
e & 116938 & 0.859 & 0.002 & 1.29 & 0.04 & 1.12 & 0.04 & 1.05 & 0.03\\ 
t & 87882 & 0.860 & 0.002 & 1.23 & 0.04 & 1.25 & 0.04 & 1.00 & 0.03\\ 
a & 77820 & 0.851 & 0.002 & 1.24 & 0.04 & 1.22 & 0.05 & 1.05 & 0.03\\ 
o & 69258 & 0.900 & 0.001 & 1.27 & 0.04 & 1.16 & 0.04 & 1.09 & 0.03\\ 
n & 65552 & 0.886 & 0.001 & 1.20 & 0.04 & 1.20 & 0.04 & 1.07 & 0.03\\ 
i & 65349 & 0.905 & 0.001 & 1.28 & 0.04 & 1.11 & 0.04 & 1.09 & 0.03\\ 
s & 64148 & 0.917 & 0.001 & 1.34 & 0.05 & 1.31 & 0.04 & 1.15 & 0.04\\ 
h & 62824 & 0.856 & 0.002 & 1.32 & 0.04 & 1.38 & 0.06 & 1.21 & 0.04\\ 
r & 52073 & 0.900 & 0.002 & 1.32 & 0.04 & 1.19 & 0.04 & 1.14 & 0.04\\ 
l & 42733 & 1.051 & 0.001 & 1.22 & 0.04 & 1.20 & 0.04 & 0.99 & 0.03\\ 
d & 38192 & 0.969 & 0.001 & 1.42 & 0.05 & 1.19 & 0.04 & 1.05 & 0.03\\ 
u & 26672 & 0.968 & 0.001 & 1.23 & 0.04 & 1.09 & 0.03 & 1.02 & 0.03\\ 
m & 23243 & 0.998 & 0.001 & 1.22 & 0.04 & 1.16 & 0.04 & 0.96 & 0.04\\ 
c & 22482 & 1.031 & 0.001 & 1.32 & 0.04 & 1.30 & 0.04 & 1.15 & 0.04\\ 
w & 22193 & 0.957 & 0.001 & 1.24 & 0.04 & 1.23 & 0.04 & 1.06 & 0.03\\ 
f & 20812 & 0.997 & 0.001 & 1.33 & 0.05 & 1.20 & 0.04 & 1.01 & 0.04\\ 
g & 20801 & 1.009 & 0.001 & 1.32 & 0.04 & 1.11 & 0.03 & 1.07 & 0.04\\ 
p & 17233 & 1.057 & 0.001 & 1.23 & 0.04 & 1.13 & 0.04 & 1.12 & 0.03\\ 
y & 16852 & 1.037 & 0.001 & 1.25 & 0.05 & 1.22 & 0.04 & 0.98 & 0.04\\ 
\_the\_ & 14404 & 1.033 & 0.002 & 1.40 & 0.04 &  &  &  & \\ 
\_of\_ & 6600 & 1.073 & 0.003 & 1.47 & 0.06 &  &  &  & \\ 
\_and\_ & 6428 & 0.962 & 0.002 & 1.23 & 0.04 &  &  &  & \\ 
\_a\_ & 4722 & 1.137 & 0.003 & 1.34 & 0.05 &  &  &  & \\ 
\_to\_ & 4619 & 1.023 & 0.003 & 1.15 & 0.04 &  &  &  & \\ 
\_in\_ & 4166 & 1.021 & 0.003 & 1.30 & 0.05 &  &  &  & \\ 
\_whale\_ & 1096 & 2.162 & 0.018 & 1.57 & 0.07 &  &  &  & \\ 
\_from\_ & 1085 & 1.143 & 0.006 & 1.15 & 0.04 &  &  &  & \\ 
\_man\_ & 476 & 1.252 & 0.007 & 1.21 & 0.04 &  &  &  & \\ 
\_them\_ & 474 & 1.214 & 0.012 & 1.16 & 0.04 &  &  &  & \\ 
\_sea\_ & 453 & 1.311 & 0.009 & 1.24 & 0.04 &  &  &  & \\ 
\_old\_ & 450 & 1.507 & 0.012 & 1.33 & 0.04 &  &  &  & \\ 
\_we\_ & 445 & 1.646 & 0.011 & 1.28 & 0.05 &  &  &  & \\ 
\_ship\_ & 438 & 1.522 & 0.012 & 1.31 & 0.04 &  &  &  & \\ 
\_ahab\_ & 436 & 3.056 & 0.021 & 1.53 & 0.06 &  &  &  & \\ 
\_ye\_ & 431 & 2.680 & 0.018 & 1.43 & 0.04 &  &  &  & \\ 
\_who\_ & 344 & 1.136 & 0.012 & 1.22 & 0.04 &  &  &  & \\ 
\_head\_ & 342 & 1.346 & 0.012 & 1.35 & 0.05 &  &  &  & \\ 
\_time\_ & 333 & 1.086 & 0.014 & 1.08 & 0.04 &  &  &  & \\ 
\_long\_ & 333 & 1.092 & 0.009 & 1.07 & 0.03 &  &  &  & \\ 
\hline
\end{tabular}
\end{table}

\begin{table}[ht]
\caption{Correlation~$\hat{\gamma}$ and burstiness $\sigma_\tau/\langle \tau \rangle$ obtained for the diferrent binary sequences in the indicated book. }
 \begin{tabular}{|c||c|c|c|c|c||c|c|c|c|}
\hline
 \multicolumn{10}{|c|}{Book: pride; N=659,408}\\
\hline
\hline
\multicolumn{2}{|c|}{} & \multicolumn{4}{|c|}{Original data  }   & \multicolumn{2}{|c|}{Shuffling  M1} & \multicolumn{2}{|c|}{Shuffling  M2} \\
\hline
sequence & $N_i$ & $\sigma_\tau/\langle \tau \rangle$ & error & $\hat{\gamma}$ & error & $\hat{\gamma}$ & error & $\hat{\gamma}$ & error\\ 
\hline
vowels & 203916 & 0.437 & 0.020 & 1.20 & 0.04 &  &  &  & \\ 
\_ & 122194 & 0.450 & 0.008 & 1.41 & 0.05 & 1.41 & 0.05 & 1.41 & 0.05\\ 
e & 69370 & 0.828 & 0.002 & 1.19 & 0.04 & 1.12 & 0.04 & 1.08 & 0.04\\ 
t & 46645 & 0.872 & 0.002 & 1.10 & 0.05 & 1.09 & 0.04 & 1.02 & 0.03\\ 
a & 41688 & 0.849 & 0.002 & 1.11 & 0.03 & 1.18 & 0.04 & 1.04 & 0.04\\ 
o & 40041 & 0.891 & 0.001 & 1.18 & 0.04 & 1.09 & 0.03 & 1.00 & 0.04\\ 
i & 37830 & 0.870 & 0.002 & 1.16 & 0.04 & 1.31 & 0.04 & 1.09 & 0.04\\ 
n & 37689 & 0.884 & 0.001 & 1.13 & 0.04 & 1.16 & 0.04 & 1.09 & 0.03\\ 
h & 34067 & 0.869 & 0.002 & 1.31 & 0.04 & 1.04 & 0.04 & 1.10 & 0.03\\ 
s & 33114 & 0.956 & 0.001 & 1.06 & 0.03 & 1.11 & 0.03 & 1.06 & 0.03\\ 
r & 32299 & 0.882 & 0.001 & 1.18 & 0.04 & 1.09 & 0.04 & 1.06 & 0.04\\ 
d & 22303 & 0.917 & 0.002 & 1.15 & 0.04 & 1.11 & 0.03 & 1.05 & 0.03\\ 
l & 21594 & 1.036 & 0.001 & 1.19 & 0.05 & 1.10 & 0.04 & 1.03 & 0.04\\ 
u & 14987 & 0.971 & 0.002 & 1.26 & 0.04 & 1.17 & 0.04 & 1.03 & 0.03\\ 
m & 14764 & 0.963 & 0.002 & 1.17 & 0.04 & 1.17 & 0.04 & 1.02 & 0.03\\ 
c & 13461 & 1.005 & 0.002 & 1.27 & 0.06 & 1.20 & 0.04 & 1.04 & 0.04\\ 
y & 12706 & 0.992 & 0.002 & 1.37 & 0.05 & 1.20 & 0.05 & 1.04 & 0.03\\ 
w & 12305 & 0.949 & 0.002 & 1.23 & 0.04 & 1.14 & 0.04 & 1.06 & 0.04\\ 
f & 11998 & 0.988 & 0.002 & 1.23 & 0.04 & 1.05 & 0.03 & 1.05 & 0.03\\ 
g & 10031 & 0.949 & 0.002 & 1.06 & 0.04 & 1.17 & 0.04 & 1.03 & 0.03\\ 
b & 9088 & 0.943 & 0.002 & 1.19 & 0.05 & 1.08 & 0.03 & 0.99 & 0.04\\ 
\_the\_ & 4331 & 1.083 & 0.003 & 1.24 & 0.04 &  &  &  & \\ 
\_to\_ & 4163 & 0.945 & 0.003 & 1.11 & 0.03 &  &  &  & \\ 
\_of\_ & 3609 & 0.974 & 0.003 & 1.21 & 0.04 &  &  &  & \\ 
\_and\_ & 3585 & 0.859 & 0.003 & 1.18 & 0.04 &  &  &  & \\ 
\_her\_ & 2225 & 1.592 & 0.015 & 1.31 & 0.04 &  &  &  & \\ 
\_i\_ & 2068 & 2.915 & 0.014 & 1.46 & 0.05 &  &  &  & \\ 
\_at\_ & 788 & 1.071 & 0.006 & 1.10 & 0.04 &  &  &  & \\ 
\_mr\_ & 786 & 1.218 & 0.007 & 1.32 & 0.06 &  &  &  & \\ 
\_they\_ & 601 & 1.459 & 0.010 & 1.26 & 0.04 &  &  &  & \\ 
\_elizabeth\_ & 597 & 1.192 & 0.027 & 1.17 & 0.06 &  &  &  & \\ 
\_or\_ & 300 & 1.026 & 0.010 & 1.00 & 0.03 &  &  &  & \\ 
\_bennet\_ & 294 & 2.047 & 0.034 & 1.37 & 0.07 &  &  &  & \\ 
\_who\_ & 284 & 1.148 & 0.010 & 1.06 & 0.03 &  &  &  & \\ 
\_miss\_ & 283 & 1.536 & 0.015 & 1.35 & 0.07 &  &  &  & \\ 
\_one\_ & 268 & 1.066 & 0.009 & 1.06 & 0.04 &  &  &  & \\ 
\_jane\_ & 264 & 1.741 & 0.016 & 1.29 & 0.06 &  &  &  & \\ 
\_bingley\_ & 257 & 3.166 & 0.019 & 1.45 & 0.08 &  &  &  & \\ 
\_we\_ & 253 & 1.546 & 0.013 & 1.26 & 0.04 &  &  &  & \\ 
\_own\_ & 183 & 1.078 & 0.015 & 1.06 & 0.04 &  &  &  & \\ 
\_lady\_ & 183 & 1.924 & 0.023 & 1.38 & 0.06 &  &  &  & \\ 
\hline
\end{tabular}
\end{table}

\begin{table}[ht]
\caption{Correlation~$\hat{\gamma}$ and burstiness $\sigma_\tau/\langle \tau \rangle$ obtained for the diferrent binary sequences in the indicated book. }
 \begin{tabular}{|c||c|c|c|c|c||c|c|c|c|}
\hline
 \multicolumn{10}{|c|}{Book: quixote; N=2,080,431}\\
\hline
\hline
\multicolumn{2}{|c|}{} & \multicolumn{4}{|c|}{Original data  }   & \multicolumn{2}{|c|}{Shuffling  M1} & \multicolumn{2}{|c|}{Shuffling  M2} \\
\hline
sequence & $N_i$ & $\sigma_\tau/\langle \tau \rangle$ & error & $\hat{\gamma}$ & error & $\hat{\gamma}$ & error & $\hat{\gamma}$ & error\\ 
\hline
vowels & 638882 & 0.430 & 0.020 & 1.26 & 0.04 &  &  &  & \\ 
\_ & 402964 & 0.415 & 0.009 & 1.34 & 0.05 & 1.34 & 0.05 & 1.34 & 0.05\\ 
e & 204300 & 0.840 & 0.002 & 1.25 & 0.04 & 1.25 & 0.04 & 1.03 & 0.03\\ 
t & 157193 & 0.867 & 0.002 & 1.26 & 0.04 & 1.25 & 0.04 & 1.03 & 0.03\\ 
a & 138706 & 0.841 & 0.002 & 1.25 & 0.04 & 1.28 & 0.04 & 1.09 & 0.03\\ 
o & 136541 & 0.881 & 0.001 & 1.24 & 0.04 & 1.29 & 0.04 & 1.03 & 0.03\\ 
h & 117821 & 0.852 & 0.002 & 1.23 & 0.04 & 1.10 & 0.04 & 1.06 & 0.03\\ 
n & 115898 & 0.866 & 0.002 & 1.23 & 0.04 & 1.17 & 0.04 & 1.05 & 0.04\\ 
i & 112746 & 0.881 & 0.001 & 1.26 & 0.04 & 1.19 & 0.04 & 1.04 & 0.03\\ 
s & 106979 & 0.935 & 0.001 & 1.28 & 0.05 & 1.27 & 0.04 & 1.06 & 0.03\\ 
r & 92501 & 0.910 & 0.001 & 1.27 & 0.04 & 1.28 & 0.04 & 1.06 & 0.03\\ 
d & 76655 & 0.929 & 0.001 & 1.34 & 0.04 & 1.22 & 0.04 & 1.03 & 0.03\\ 
l & 62107 & 1.108 & 0.002 & 1.24 & 0.04 & 1.28 & 0.05 & 1.01 & 0.03\\ 
u & 46589 & 0.949 & 0.001 & 1.23 & 0.04 & 1.17 & 0.04 & 1.00 & 0.04\\ 
m & 42945 & 0.992 & 0.001 & 1.21 & 0.04 & 1.25 & 0.04 & 1.04 & 0.05\\ 
f & 38552 & 0.977 & 0.001 & 1.24 & 0.04 & 1.24 & 0.04 & 1.05 & 0.03\\ 
w & 38209 & 0.986 & 0.001 & 1.29 & 0.04 & 1.21 & 0.04 & 1.05 & 0.03\\ 
c & 37602 & 0.984 & 0.001 & 1.21 & 0.04 & 1.26 & 0.04 & 1.08 & 0.03\\ 
g & 31927 & 0.988 & 0.001 & 1.22 & 0.04 & 1.23 & 0.04 & 1.03 & 0.03\\ 
y & 31053 & 1.048 & 0.001 & 1.25 & 0.04 & 1.26 & 0.04 & 1.05 & 0.04\\ 
p & 23880 & 1.069 & 0.001 & 1.23 & 0.04 & 1.26 & 0.04 & 1.11 & 0.03\\ 
\_the\_ & 20652 & 1.050 & 0.002 & 1.36 & 0.04 &  &  &  & \\ 
\_and\_ & 16835 & 0.908 & 0.002 & 1.22 & 0.05 &  &  &  & \\ 
\_to\_ & 13184 & 1.031 & 0.002 & 1.26 & 0.04 &  &  &  & \\ 
\_of\_ & 12173 & 1.033 & 0.002 & 1.26 & 0.04 &  &  &  & \\ 
\_that\_ & 7515 & 1.023 & 0.002 & 1.21 & 0.04 &  &  &  & \\ 
\_in\_ & 6716 & 1.023 & 0.002 & 1.11 & 0.03 &  &  &  & \\ 
\_by\_ & 2069 & 1.042 & 0.004 & 1.09 & 0.03 &  &  &  & \\ 
\_sancho\_ & 2063 & 3.762 & 0.025 & 1.63 & 0.05 &  &  &  & \\ 
\_or\_ & 2048 & 1.154 & 0.004 & 1.15 & 0.04 &  &  &  & \\ 
\_quixote\_ & 2002 & 3.214 & 0.016 & 1.55 & 0.06 &  &  &  & \\ 
\_other\_ & 609 & 1.072 & 0.008 & 1.11 & 0.04 &  &  &  & \\ 
\_knight\_ & 606 & 2.175 & 0.016 & 1.43 & 0.05 &  &  &  & \\ 
\_take\_ & 546 & 1.195 & 0.008 & 1.14 & 0.04 &  &  &  & \\ 
\_master\_ & 545 & 1.720 & 0.013 & 1.38 & 0.04 &  &  &  & \\ 
\_thy\_ & 510 & 2.252 & 0.017 & 1.35 & 0.05 &  &  &  & \\ 
\_senor\_ & 509 & 1.632 & 0.009 & 1.28 & 0.04 &  &  &  & \\ 
\_worship\_ & 470 & 2.337 & 0.012 & 1.37 & 0.04 &  &  &  & \\ 
\_here\_ & 467 & 1.237 & 0.007 & 1.14 & 0.04 &  &  &  & \\ 
\_god\_ & 467 & 1.169 & 0.011 & 1.12 & 0.04 &  &  &  & \\ 
\_way\_ & 466 & 1.056 & 0.006 & 1.07 & 0.04 &  &  &  & \\ 
\hline
\end{tabular}
\end{table}

\begin{table}[ht]
\caption{Correlation~$\hat{\gamma}$ and burstiness $\sigma_\tau/\langle \tau \rangle$ obtained for the diferrent binary sequences in the indicated book. }
 \begin{tabular}{|c||c|c|c|c|c||c|c|c|c|}
\hline
 \multicolumn{10}{|c|}{Book: sawyer; N=369,222 }\\
\hline
\hline
\multicolumn{2}{|c|}{} & \multicolumn{4}{|c|}{Original data  }   & \multicolumn{2}{|c|}{Shuffling  M1} & \multicolumn{2}{|c|}{Shuffling  M2} \\
\hline
sequence & $N_i$ & $\sigma_\tau/\langle \tau \rangle$ & error & $\hat{\gamma}$ & error & $\hat{\gamma}$ & error & $\hat{\gamma}$ & error\\ 
\hline
vowels & 110026 & 0.432 & 0.020 & 1.23 & 0.04 &  &  &  & \\ 
\_ & 71180 & 0.402 & 0.010 & 1.50 & 0.05 & 1.50 & 0.05 & 1.50 & 0.05\\ 
e & 35603 & 0.864 & 0.002 & 1.30 & 0.04 & 1.05 & 0.05 & 0.99 & 0.04\\ 
t & 28825 & 0.858 & 0.002 & 1.16 & 0.04 & 1.23 & 0.04 & 0.99 & 0.04\\ 
a & 23478 & 0.858 & 0.002 & 1.17 & 0.04 & 1.03 & 0.03 & 1.13 & 0.04\\ 
o & 23192 & 0.898 & 0.001 & 1.22 & 0.04 & 1.06 & 0.03 & 0.98 & 0.04\\ 
n & 20146 & 0.866 & 0.002 & 1.12 & 0.04 & 1.23 & 0.04 & 1.07 & 0.03\\ 
h & 19565 & 0.861 & 0.002 & 1.18 & 0.04 & 1.11 & 0.04 & 1.07 & 0.04\\ 
i & 18811 & 0.910 & 0.002 & 1.16 & 0.05 & 1.23 & 0.04 & 1.05 & 0.03\\ 
s & 17716 & 0.951 & 0.001 & 1.19 & 0.04 & 1.13 & 0.04 & 1.08 & 0.03\\ 
r & 15247 & 0.917 & 0.002 & 1.30 & 0.05 & 1.13 & 0.04 & 1.10 & 0.05\\ 
d & 14850 & 0.950 & 0.002 & 1.20 & 0.04 & 1.20 & 0.04 & 1.01 & 0.03\\ 
l & 12136 & 1.086 & 0.002 & 1.17 & 0.04 & 1.14 & 0.04 & 1.06 & 0.03\\ 
u & 8942 & 0.949 & 0.002 & 1.18 & 0.04 & 1.07 & 0.04 & 1.06 & 0.03\\ 
w & 8042 & 0.949 & 0.002 & 1.13 & 0.03 & 1.18 & 0.04 & 1.12 & 0.04\\ 
m & 7135 & 0.977 & 0.002 & 1.22 & 0.04 & 1.18 & 0.04 & 1.02 & 0.03\\ 
y & 6725 & 1.043 & 0.002 & 1.36 & 0.04 & 1.04 & 0.03 & 1.00 & 0.04\\ 
g & 6606 & 1.041 & 0.002 & 1.16 & 0.04 & 1.15 & 0.05 & 1.06 & 0.03\\ 
c & 6497 & 1.030 & 0.003 & 1.23 & 0.05 & 1.09 & 0.05 & 1.16 & 0.04\\ 
f & 6004 & 1.047 & 0.003 & 1.22 & 0.04 & 1.11 & 0.03 & 1.02 & 0.03\\ 
b & 4958 & 0.959 & 0.003 & 1.10 & 0.04 & 1.25 & 0.04 & 1.02 & 0.03\\ 
\_the\_ & 3703 & 1.154 & 0.004 & 1.35 & 0.04 &  &  &  & \\ 
\_and\_ & 3105 & 1.008 & 0.003 & 1.21 & 0.04 &  &  &  & \\ 
\_a\_ & 1863 & 1.085 & 0.005 & 1.20 & 0.04 &  &  &  & \\ 
\_to\_ & 1727 & 1.054 & 0.004 & 1.14 & 0.03 &  &  &  & \\ 
\_of\_ & 1436 & 1.127 & 0.005 & 1.21 & 0.04 &  &  &  & \\ 
\_he\_ & 1197 & 1.770 & 0.015 & 1.40 & 0.04 &  &  &  & \\ 
\_tom\_ & 689 & 1.740 & 0.014 & 1.39 & 0.06 &  &  &  & \\ 
\_with\_ & 647 & 1.068 & 0.008 & 1.15 & 0.04 &  &  &  & \\ 
\_if\_ & 237 & 1.404 & 0.011 & 1.23 & 0.04 &  &  &  & \\ 
\_huck\_ & 223 & 3.228 & 0.024 & 1.46 & 0.07 &  &  &  & \\ 
\_boys\_ & 155 & 1.767 & 0.019 & 1.24 & 0.06 &  &  &  & \\ 
\_did\_ & 150 & 1.336 & 0.018 & 1.22 & 0.04 &  &  &  & \\ 
\_joe\_ & 133 & 2.248 & 0.051 & 1.38 & 0.06 &  &  &  & \\ 
\_never\_ & 131 & 1.185 & 0.017 & 1.14 & 0.04 &  &  &  & \\ 
\_boy\_ & 122 & 1.788 & 0.054 & 1.29 & 0.06 &  &  &  & \\ 
\_back\_ & 121 & 0.968 & 0.015 & 1.04 & 0.03 &  &  &  & \\ 
\_off\_ & 99 & 1.335 & 0.019 & 1.12 & 0.04 &  &  &  & \\ 
\_night\_ & 98 & 2.025 & 0.057 & 1.29 & 0.04 &  &  &  & \\ 
\_other\_ & 96 & 1.145 & 0.019 & 1.12 & 0.03 &  &  &  & \\ 
\_becky\_ & 96 & 2.701 & 0.036 & 1.55 & 0.10 &  &  &  & \\ 
\hline
\end{tabular}
\end{table}

\begin{table}[ht]
\caption{Correlation~$\hat{\gamma}$ and burstiness $\sigma_\tau/\langle \tau \rangle$ obtained for the diferrent binary sequences in the
  indicated book. }
 \begin{tabular}{|c||c|c|c|c|c||c|c|c|c|}
\hline
 \multicolumn{10}{|c|}{Book: ulysses; N=1,453,586 }\\
\hline
\hline
\multicolumn{2}{|c|}{} & \multicolumn{4}{|c|}{Original data  }   & \multicolumn{2}{|c|}{Shuffling  M1} & \multicolumn{2}{|c|}{Shuffling  M2} \\
\hline
sequence & $N_i$ & $\sigma_\tau/\langle \tau \rangle$ & error & $\hat{\gamma}$ & error & $\hat{\gamma}$ & error & $\hat{\gamma}$ & error\\ 
\hline
vowels & 440676 & 0.456 & 0.020 & 1.61 & 0.06 &  &  &  & \\ 
\_ & 265304 & 0.436 & 0.009 & 1.78 & 0.06 & 1.78 & 0.06 & 1.78 & 0.06\\ 
e & 141465 & 0.855 & 0.002 & 1.28 & 0.04 & 1.30 & 0.06 & 1.11 & 0.03\\ 
t & 100183 & 0.904 & 0.001 & 1.54 & 0.07 & 1.37 & 0.06 & 1.05 & 0.03\\ 
a & 93129 & 0.877 & 0.001 & 1.32 & 0.05 & 1.25 & 0.06 & 1.06 & 0.03\\ 
o & 91403 & 0.930 & 0.001 & 1.19 & 0.04 & 1.35 & 0.05 & 1.10 & 0.04\\ 
i & 81407 & 0.914 & 0.001 & 1.44 & 0.07 & 1.33 & 0.06 & 1.21 & 0.05\\ 
n & 80138 & 0.897 & 0.001 & 1.34 & 0.05 & 1.27 & 0.04 & 1.26 & 0.06\\ 
s & 76915 & 0.950 & 0.001 & 1.40 & 0.06 & 1.31 & 0.06 & 1.23 & 0.06\\ 
h & 72550 & 0.906 & 0.002 & 1.61 & 0.06 & 1.23 & 0.05 & 1.44 & 0.08\\ 
r & 69852 & 0.918 & 0.001 & 1.49 & 0.06 & 1.22 & 0.04 & 1.30 & 0.06\\ 
l & 55052 & 1.074 & 0.001 & 1.41 & 0.06 & 1.39 & 0.06 & 1.19 & 0.05\\ 
d & 49093 & 0.980 & 0.001 & 1.44 & 0.05 & 1.17 & 0.04 & 1.04 & 0.04\\ 
u & 33272 & 0.982 & 0.001 & 1.25 & 0.04 & 1.36 & 0.06 & 1.16 & 0.04\\ 
m & 31535 & 1.025 & 0.001 & 1.29 & 0.05 & 1.35 & 0.05 & 1.05 & 0.03\\ 
c & 29894 & 1.072 & 0.001 & 1.62 & 0.07 & 1.31 & 0.04 & 1.36 & 0.07\\ 
g & 27791 & 1.031 & 0.001 & 1.36 & 0.06 & 1.24 & 0.05 & 1.19 & 0.04\\ 
f & 26638 & 1.025 & 0.001 & 1.30 & 0.05 & 1.22 & 0.04 & 1.12 & 0.03\\ 
w & 26164 & 1.056 & 0.001 & 1.53 & 0.07 & 1.32 & 0.05 & 1.31 & 0.06\\ 
y & 24251 & 1.032 & 0.001 & 1.36 & 0.05 & 1.15 & 0.03 & 1.01 & 0.03\\ 
p & 22440 & 1.124 & 0.002 & 1.46 & 0.06 & 1.27 & 0.05 & 1.25 & 0.05\\ 
\_the\_ & 14952 & 1.071 & 0.002 & 1.44 & 0.06 &  &  &  & \\ 
\_of\_ & 8141 & 1.121 & 0.003 & 1.63 & 0.07 &  &  &  & \\ 
\_and\_ & 7217 & 1.167 & 0.003 & 1.53 & 0.05 &  &  &  & \\ 
\_a\_ & 6518 & 1.144 & 0.003 & 1.24 & 0.04 &  &  &  & \\ 
\_to\_ & 4963 & 1.157 & 0.003 & 1.38 & 0.07 &  &  &  & \\ 
\_in\_ & 4946 & 1.002 & 0.002 & 1.16 & 0.04 &  &  &  & \\ 
\_were\_ & 510 & 1.461 & 0.013 & 1.27 & 0.04 &  &  &  & \\ 
\_stephen\_ & 505 & 4.955 & 0.099 & 1.64 & 0.06 &  &  &  & \\ 
\_we\_ & 425 & 2.427 & 0.085 & 1.25 & 0.04 &  &  &  & \\ 
\_man\_ & 415 & 1.388 & 0.019 & 1.16 & 0.04 &  &  &  & \\ 
\_into\_ & 330 & 1.179 & 0.011 & 1.14 & 0.04 &  &  &  & \\ 
\_eyes\_ & 329 & 1.921 & 0.013 & 1.21 & 0.04 &  &  &  & \\ 
\_where\_ & 310 & 1.214 & 0.014 & 1.11 & 0.03 &  &  &  & \\ 
\_hand\_ & 308 & 1.295 & 0.017 & 1.18 & 0.04 &  &  &  & \\ 
\_street\_ & 293 & 1.394 & 0.013 & 1.21 & 0.04 &  &  &  & \\ 
\_our\_ & 291 & 1.556 & 0.018 & 1.23 & 0.04 &  &  &  & \\ 
\_first\_ & 278 & 1.306 & 0.011 & 1.19 & 0.04 &  &  &  & \\ 
\_father\_ & 277 & 1.631 & 0.013 & 1.62 & 0.05 &  &  &  & \\ 
\_day\_ & 250 & 1.131 & 0.012 & 1.10 & 0.03 &  &  &  & \\ 
\_just\_ & 249 & 2.014 & 0.012 & 1.20 & 0.04 &  &  &  & \\ 
\hline
\end{tabular}
\end{table}

\begin{table}[ht]
\caption{Correlation~$\hat{\gamma}$ and burstiness $\sigma_\tau/\langle \tau \rangle$ obtained for the diferrent binary sequences in the indicated book. }
 \begin{tabular}{|c||c|c|c|c|c||c|c|c|c|}
\hline
 \multicolumn{10}{|c|}{Book: wrnpc; N=3,082,079} \\
\hline
\hline
\multicolumn{2}{|c|}{} & \multicolumn{4}{|c|}{Original data  }   & \multicolumn{2}{|c|}{Shuffling  M1} & \multicolumn{2}{|c|}{Shuffling  M2} \\
\hline
sequence & $N_i$ & $\sigma_\tau/\langle \tau \rangle$ & error & $\hat{\gamma}$ & error & $\hat{\gamma}$ & error & $\hat{\gamma}$ & error\\ 
\hline
vowels & 945097 & 0.430 & 0.020 & 1.55 & 0.05 &  &  &  & \\ 
\_ & 565161 & 0.426 & 0.009 & 1.50 & 0.05 & 1.50 & 0.05 & 1.50 & 0.05\\ 
e & 312626 & 0.834 & 0.002 & 1.39 & 0.05 & 1.35 & 0.04 & 1.05 & 0.03\\ 
t & 224180 & 0.886 & 0.001 & 1.40 & 0.04 & 1.27 & 0.04 & 1.09 & 0.03\\ 
a & 204154 & 0.869 & 0.002 & 1.42 & 0.05 & 1.18 & 0.04 & 1.05 & 0.03\\ 
o & 191126 & 0.904 & 0.001 & 1.45 & 0.05 & 1.40 & 0.04 & 1.04 & 0.04\\ 
n & 182910 & 0.860 & 0.002 & 1.28 & 0.04 & 1.43 & 0.05 & 1.17 & 0.04\\ 
i & 172403 & 0.894 & 0.001 & 1.48 & 0.05 & 1.46 & 0.05 & 1.21 & 0.04\\ 
h & 166290 & 0.852 & 0.002 & 1.50 & 0.05 & 1.26 & 0.04 & 1.28 & 0.05\\ 
s & 161889 & 0.955 & 0.001 & 1.32 & 0.04 & 1.47 & 0.05 & 1.04 & 0.03\\ 
r & 146667 & 0.919 & 0.001 & 1.38 & 0.04 & 1.34 & 0.04 & 1.10 & 0.03\\ 
d & 117632 & 0.923 & 0.001 & 1.48 & 0.05 & 1.33 & 0.04 & 1.05 & 0.03\\ 
l & 95888 & 1.064 & 0.001 & 1.26 & 0.04 & 1.30 & 0.04 & 1.04 & 0.04\\ 
u & 64788 & 0.971 & 0.001 & 1.26 & 0.04 & 1.26 & 0.04 & 1.04 & 0.03\\ 
m & 61162 & 1.018 & 0.001 & 1.30 & 0.04 & 1.26 & 0.04 & 0.98 & 0.04\\ 
c & 60576 & 1.009 & 0.001 & 1.54 & 0.05 & 1.39 & 0.04 & 1.17 & 0.04\\ 
w & 58852 & 0.978 & 0.001 & 1.28 & 0.04 & 1.29 & 0.04 & 1.24 & 0.05\\ 
f & 54419 & 1.064 & 0.001 & 1.49 & 0.05 & 1.23 & 0.04 & 1.05 & 0.03\\ 
g & 50819 & 1.014 & 0.001 & 1.49 & 0.05 & 1.37 & 0.04 & 1.08 & 0.04\\ 
y & 45847 & 1.035 & 0.001 & 1.36 & 0.04 & 1.34 & 0.04 & 1.01 & 0.04\\ 
p & 44680 & 1.080 & 0.001 & 1.48 & 0.05 & 1.34 & 0.04 & 1.12 & 0.03\\ 
\_the\_ & 34495 & 1.128 & 0.002 & 1.59 & 0.05 &  &  &  & \\ 
\_and\_ & 22217 & 0.874 & 0.002 & 1.22 & 0.04 &  &  &  & \\ 
\_to\_ & 16640 & 1.056 & 0.001 & 1.24 & 0.04 &  &  &  & \\ 
\_of\_ & 14864 & 1.168 & 0.002 & 1.59 & 0.05 &  &  &  & \\ 
\_a\_ & 10525 & 1.119 & 0.002 & 1.21 & 0.04 &  &  &  & \\ 
\_he\_ & 9860 & 1.893 & 0.006 & 1.44 & 0.06 &  &  &  & \\ 
\_so\_ & 1900 & 1.180 & 0.005 & 1.14 & 0.03 &  &  &  & \\ 
\_prince\_ & 1890 & 3.862 & 0.026 & 1.68 & 0.06 &  &  &  & \\ 
\_pierre\_ & 1796 & 6.563 & 0.042 & 1.78 & 0.06 &  &  &  & \\ 
\_an\_ & 1625 & 1.131 & 0.004 & 1.17 & 0.04 &  &  &  & \\ 
\_could\_ & 1115 & 1.285 & 0.005 & 1.15 & 0.05 &  &  &  & \\ 
\_natasha\_ & 1098 & 6.334 & 0.036 & 1.74 & 0.06 &  &  &  & \\ 
\_man\_ & 1081 & 1.407 & 0.007 & 1.33 & 0.04 &  &  &  & \\ 
\_will\_ & 1066 & 1.530 & 0.011 & 1.36 & 0.04 &  &  &  & \\ 
\_andrew\_ & 1047 & 4.321 & 0.021 & 1.71 & 0.06 &  &  &  & \\ 
\_do\_ & 1037 & 1.273 & 0.010 & 1.16 & 0.04 &  &  &  & \\ 
\_time\_ & 926 & 1.100 & 0.008 & 1.13 & 0.03 &  &  &  & \\ 
\_princess\_ & 915 & 5.431 & 0.033 & 1.72 & 0.06 &  &  &  & \\ 
\_face\_ & 893 & 1.445 & 0.007 & 1.25 & 0.04 &  &  &  & \\ 
\_french\_ & 879 & 2.287 & 0.029 & 1.53 & 0.05 &  &  &  & \\ 
\hline
\end{tabular}
\end{table}


\begin{thebibliography}{99}

\bibitem{schenkel} Schenkel A, Zhang J, Zhang Y (1993)
Long range correlation in human writings.
{\it Fractals} 1:47-55.

\bibitem{eckmann1} Alvarez-Lacalle E, Dorow B, Eckmann JP, Moses E, (2006)
Hierarchical structures induce long-range dynamical correlations in written texts.
{\it Proc Natl Acad Sci USA} 103:7956-7961.


\bibitem{vossMusic} Voss R, Clarke J (1975)
`1/f noise' in music and speech.
{\it Nature} 258:317-318.

\bibitem{gilden} Gilden D, Thornton T, Mallon M (1995)
1/f noise in human cognition.
{\it Science} 267:1837-1839.

\bibitem{bunde} Muchnik L, Havlin S, Bunde A, Stanley HE (2005)
Scaling and memory in volatility return intervals in financial markets.
{\it Proc Natl Acad Sci USA} 102:9424-9428.

\bibitem{rybski} Rybski D, Buldyrev SV, Havlin S, Liljeros F, Makse HA (2009)
Scaling laws of human interaction activity.
{\it Proc Natl Acad Sci} 106:12640-12645.

\bibitem{kello}  Kello CT, Brown GDA, Ferrer-i-Cancho R, Holden JG, Linkenkaer-Hansen K, Rhodes T, Van Orden GC (2010)
Scaling laws in cognitive sciences.
{\it Trends Cogn Sci} 14:223-232

\bibitem{press}
Press WH (1978) Flicker Noises in Astronomy and Elsewhere. 
{\it Comments on Astrophysics} 7:103 

\bibitem{li} Li W, Kaneko K (1992)
Long-range correlation and partial $1/f^\alpha$ spectrum in a noncoding DNA sequence.
{\it Europhys Lett}  17:655-660.

\bibitem{peng} Peng CK, Buldyrev S, Goldberger A, Havlin S, Sciortino F, Simons M, and Stanley HE (1992)
Long-Range Correlations in Nucleotide Sequences.
{\it Nature} 356: 168-171.

\bibitem{voss}
Voss RF (1992)
Evolution of long-range fractal correlations and $1/f$ noise in DNA base sequences.
{\it Phys Rev Lett} 68:3805-3808.

\bibitem{manni} C.D. Manning, H. Sch\"utze (1999)
{\it Foundations of Statistical Natural Language Processing}, (The MIT Press, Cambridge, Massachusetts, USA).


\bibitem{stamata} Stamatatos E (2009)
A survey of modern authorship attribution methods.
{\it Journal of the American Society for Information Science and Technology} 60:538-556.


\bibitem{ober} Oberlander J and Brew C (2000)
Stochastic text generation.
{\it Phil Trans R Soc Lond A} 358:1373-1387.


\bibitem{usatenko} O Usatenko, V Yampolskii (2003)
Binary N-Step Markov Chains and Long-Range Correlated Systems.
{\it Phys Rev Lett} 90:110601.

\bibitem{amit} Amit M, Shmerler Y, Eisenberg E, Abraham M, Shnerb N (1994)
Language and codification dependence of long-range correlations in texts.
{\it Fractals} 2:7-13

\bibitem{ebeling1} Ebeling W, Neiman A (1995)
Long-range correlations between letters and sentences in texts.
{\it Physica A} 215:233-241.


\bibitem{ebeling3} Ebeling W, P\"oschel T (1994)
Entropy and long-range correlations in literary English.
{\it Europhys Lett} 26:241-246.

\bibitem{allegrini} Allegrini P, Grigolini P, Palatella L (2004)
Intermittency and scale-free networks: a dynamical model for human language complexity.
{\it Chaos, Solitons and Fractals} 20:95-105.

\bibitem{melnyk2005}
Melnyk SS, Usatenko OV, and Yampolskii VA (2005)
Competition between two kinds of correlations in literary texts.
{\it Phys Rev E} 72:026140.


\bibitem{herrera}  Herrera JP, Pury PA (2008)
Statistical keyword detection in literary corpora.
{\it Eur Phys J B} 63:135-146.

\bibitem{montemurro}
Montemurro MA, Zanette D (2010)
Towards the quantification of the semantic information encoded in  written language.
{\it Adv Comp Syst} 13:135-153.



\bibitem{cover} Cover TM, Thomas JA (2006) \textit{Elements of Information Theory} (Wiley Series in Telecommunications and Signal
Processing)


\bibitem {HerGro95} Herzel H, Gro\ss e I (1995) 
Measuring correlations in symbol sequences. 
{\it Physica A: Statistical Mechanics and its Applications}, 216:518-542

\bibitem{grassberger} Grassberger P  (1989)
Estimating the information content of symbol sequences and efficient codes.
{\it IEEE Transactions on Information Theory}, 35:669-675.

\bibitem{kokol} Kokol P, Podgorelec V (2000)
Complexity And Human Writings .
{\it Complexity} 7:1-6.


\bibitem{kanter} Kanter I, Kessler DA (1995)
Markov processes: linguistics and Zipf's Law.
{\it Phys Rev Lett} 74:4559-4562.

\bibitem{monte} Montemurro MA, Pury PA (2002)
Long-range fractal correlations in literary corpora.
{\it Fractals} 10:451-461.

\bibitem{trefan} Tref\'an G, Floriani E, West BJ and Grigolini P, (1994)
Dynamical approach to anomalous diffusion: response of Levy processes to a perturbation.
{\it Phys Rev E} 50:2564-2579.


\bibitem{Cox} 
Cox DR, Lewis PAW (1978)
\textit{The statistical analysis of series of events} (Chapman and Hall, London).

\bibitem{Benj} B. Lindner (2006)
Superposition of many independent spike trains is generally not a Poisson process.
{\it Phys Rev E} 73:022901.

\bibitem{allegrini2} 
Allegrini P, Menicucci D, Bedini R, Gemignani A, Paradisi P (2010)
Complex intermittency blurred by noise: Theory and application to neural dynamics.
{\it Phys Rev E} 82:015103.




\bibitem{goh}
Goh K-I, Barabasi A-L,
Burstiness and memory in complex systems.
{\it Europhys Lett}  81: 48002.


\bibitem{ortuno} Ortuno M, Carpena P, Bernaola-Galvan P, Munoz E, Somoza AM (2002)
Keyword detection in natural languages and DNA.
{\it Europhys Lett} 57:759-764.

\bibitem{altmann} Altmann EG, Pierrehumbert JB, Motter AE (2009)
Beyond word frequency: Bursts, lulls, and scaling in the temporal distributions of words.
{\it PLoS ONE} 4:e7678.

\bibitem{doxas} Doxas I, Dennis S, Oliver WL (2009) 
The dimensionality of discourse.
{\it Proc Natl Acad Science USA} 107:4866-4871.

\bibitem{saussaure} Saussure F de (1983) 
Course in General Linguistics, Eds. Charles Bally and Albert Sechehaye. (Trans. Roy Harris. La Salle, Illinois)


\bibitem{badalamenti} Badalamenti AF (2001)
Speech Parts as Poisson Processes.
{\it Journal of Psycholinguistic Research} 30:31.


\bibitem{ScEbHe96}Schmitt AO, Ebeling W, Herzel H (1996)
The modular structure of informational sequences.
{\it  Biosystems} 37:199Ð210.


\end{thebibliography}

\begin{thebibliography}{99}

\bibitem{landauer} Landauer TK, Foltz P, Laham D (1998) 
Introduction to latent semantic analysis.
{\it Discourse Process} 25:259-284.


\bibitem{eckmann1} Alvarez-Lacalle E, Dorow B, Eckmann JP, Moses E, (2006)
Hierarchical structures induce long-range dynamical correlations in written texts.
{\it Proc Natl Acad Sci USA} 103:7956-7961.


\bibitem{Mantegna} Mantegna RN and Stanley EH (1994)
Stochastic Process with Ultraslow Convergence to a Gaussian: The Truncated L\'evy Flight
{\it  Phys. Rev. Lett} 73:2946–2949. 

\bibitem{Shlesinger} Shlesinger MF   (1995)
Comment on “Stochastic Process with Ultraslow Convergence to a Gaussian: The Truncated L\'evy Flight”
{\it Phys. Rev. Lett.} 74: 4959–4959

\bibitem{DelC} del-Castillo-Negrete D (2009)
Truncation effects in superdiffusive front propagation with L\'evy flights 
{\it Phys. Rev. E} 79: 031120.

\bibitem{ebeling3} Ebeling W, P\"oschel T (1994)
Entropy and long-range correlations in literary English.
{\it Europhys Lett} 26:241-246.

\bibitem{monte} Montemurro MA, Pury PA (2002)
Long-range fractal correlations in literary corpora.
{\it Fractals} 10:451-461.


\bibitem{ebeling1} Ebeling W, Neiman A (1995)
Long-range correlations between letters and sentences in texts.
{\it Physica A} 215:233-241.


\end{thebibliography}
\end{document}